\documentclass[a4paper,11pt]{article}
\pdfoutput=1 

\usepackage{jheppub} 
\usepackage{bm,amssymb,slashed,graphicx,multirow,soul,mathtools,xspace,array,tikz}
\usepackage[compat=1.1.0]{tikz-feynman}   
\usepackage{float}                   
\allowdisplaybreaks
\usepackage{ bbold }
\usepackage{subfigure}
\newcommand{\op}{{\cal O}}

\newcommand{\C}{{\cal C}}

\newcommand{\todo}[1]{{\color{red} \ifmmode\else[todo]\fi #1}}

\newcommand{\lp}{\left(}
\newcommand{\rp}{\right)}

\newcommand{\g}{\gamma}

\newcommand{\beq}{\begin{equation} }
\newcommand{\eeq}{\end{equation}} 
\newcommand{\bi}{\begin{itemize} }
\newcommand{\ei}{\end{itemize} }

\definecolor{Red}{rgb}{1.,0.,0.}
\definecolor{Grn}{rgb}{0.,0.75,0.}
\definecolor{Blu}{rgb}{0.,0.,1.}

\usepackage[T1]{fontenc} 

\usepackage{amsmath,amssymb,epsfig,ulem,color,slashed}
\allowdisplaybreaks  

\title{\boldmath Interplay of New Physics Effects in \boldmath $(g-2)_\ell$ and $h \to \ell^+ \ell^-$ -- Lessons from SMEFT}

\author[1,2]{Svjetlana Fajfer,}
\author[1,2]{Jernej F. Kamenik,}
\author[1]{M. Tammaro,}

\affiliation[1]{Jozef Stefan Institute, Jamova 39, Ljubljana, Slovenia}
\affiliation[2]{Faculty of Mathematics and Physics, University of Ljubljana, Jadranska 19, 1000 Ljubljana, Slovenia}

\emailAdd{svjetlana.fajfer@ijs.si}
\emailAdd{jernej.kamenik@cern.ch}
\emailAdd{Michele.Tammaro@ijs.si}

\abstract{We explore the interplay of New Physics (NP) effects in $(g-2)_\ell$ and $h \to \ell^+ \ell^-$ within the Standard Model Effective Field Theory (SMEFT) framework, including one-loop Renormalization Group (RG) evolution of the Wilson coefficients as well as matching to the observables below the electroweak symmetry breaking scale. We include both the leading dimension six chirality flipping operators including a Higgs and $SU(2)_L$ gauge bosons as well as four-fermion scalar and tensor operators, forming a closed operator set under the SMEFT RG equations. We compare present and future experimental sensitivity to different representative benchmark scenarios. We also consider two simple UV completions, a Two Higgs Doublet Model and a single scalar LeptoQuark extension of the SM, and show how tree level matching to SMEFT followed by the one-loop RG evolution down to the electroweak scale can reproduce with high accuracy the $(g-2)_\ell$ and $h \to \ell^+ \ell^-$ contributions obtained by the complete one- and even two-loop calculations in the full models.}

\begin{document} 

\maketitle

\flushbottom

\section{Introduction}
\label{sec:intro}
The anomalous magnetic moments of leptons have historically played an important role in the ongoing quest to uncover and understand fundamental physics at very short distances~\cite{Schwinger:1948iu}.  
In the past two decades, especially the anomalous magnetic moment of the muon $(g-2)_{\mu}\equiv a_\mu$ has been an intriguing venue of New Physics (NP) searches due to the persistent disagreement between its measured and theoretically predicted value within the Standard Model (SM).  The currently most precise measurement by the Brookhaven experiment E281~\cite{Bennett:2006fi} is expected to be soon updated by the Muon g-2~\cite{Grange:2015fou} experiment at Fermilab.
On theory side, a lot of effort in recent years has been put into re-evaluating hadronic corrections to the muon-photon vertex \cite{Aoyama:2020ynm,Jegerlehner:2009ry,Keshavarzi:2018mgv,Davier:2019can,Keshavarzi_2020}.
Nonetheless, the current $a_\mu$ prediction within the SM\footnote{Here the theoretical value $a_\mu^{\rm SM}$ does not include the results from the recent BMWc Lattice QCD study~\cite{Borsanyi:2020mff}. We discuss the impact of this intriguing new result on BSM phenomenology in Sec.~\ref{subsec:pheno:muon}.}  deviates from the experimental result at the $3.3\sigma$ level.
Intriguingly, recent re-evaluations of the fine-structure constant have shifted the theoretical SM prediction for $a_e$~\cite{Aoyama:2017uqe} which is completely dominated by higher order QED effects. In particular, two measurements performed on Cs~\cite{Parker:2018vye} and Rb~\cite{Morel:2020dww} atoms yield results which are inconsistent with each other at more than $5\sigma$, and are also both in tension with the measured value of $a_e$ at the $2.4\sigma$ and $1.6\sigma$ level, respectively.

On the other hand, Higgs boson decays to charged fermions are key observables at high energy colliders. They represent most direct probes of the Yukawa interactions, which can be affected by BSM degrees of freedom mixing with either the Higgs boson or the SM fermions.  While Higgs interactions with the third generation SM fermions have been established soon after its discovery, the first direct measurement of $h \to \mu^+ \mu^-$ has only been obtained recently by both ATLAS~\cite{Aad:2020xfq} and CMS~\cite{Sirunyan:2020two}. The current observation at around $3\sigma$ significance leaves ample room for possible sizable NP effects.

While phenomenologically very distinct, anomalous magnetic moments and the corresponding fermionic Higgs decays are naturally correlated in BSM models. The chiral structure of the interaction is the same, as it involves a right- and a left-handed fermion. The Lorentz structure is of course different: the Higgs decay is mediated by a purely scalar vertex, while the $g-2$ arises from a tensor contraction with the photon field $F_{\mu\nu}$. However, as we will show, tensor operators arise generically from scalars (and viceversa), since they mix under the electroweak (EW) renormalization group (RG) evolution at one-loop.   

In this paper, we describe imprints of possible NP above the EW scale on $a_\ell$ and $h\to \ell^+ \ell^-$ in a model-independent fashion, adopting the framework of Standard Model Effective Field Theory (SMEFT)\footnote{Studies of explicit UV models correlating $a_\ell$ and $h\to \ell^+ \ell^-$ observables can be found e.g in Refs~\cite{kowalska2020minimal,zhang2020125}}. We focus on the leading dimension six lepton chirality-flipping operators including a Higgs and $SU_L(2)$ gauge bosons respectively. While these operators are sufficient for a tree-level analysis, they do not form a closed set under the SM RGE. For a consistent treatment, it is thus necessary to include four-fermion scalar and tensor operators, which in turn play a crucial role in the analysis of NP effects in both observables. As already noted in Ref. \cite{Feruglio:2018fxo}, including tensor operators in the RGE leads to the so-called chiral enhancement effects in the anomalous magnetic moments. At the same time, the scalar operators can generate sizable corrections to fermionic Higgs decays. In the following we perform a complete analysis of NP effects in $g-2$ and Higgs decays within SMEFT by running the relevant operators from the NP matching scale to the weak scale and evaluate their contributions to the relevant observables. A similar approach was recently taken in Ref.~\cite{aebischer2021effective}, where the focus was instead on the running of the low energy effective theory (LEFT) for $a_\ell$.

We further explore the role of four fermion operators by matching SMEFT to an explicit UV NP model. Suitable scenarios include models of LeptoQuarks~\cite{Dorsner:2019itg,Crivellin:2020mjs,Crivellin:2020tsz} as well as models with multiple Higgs fields. Indeed, we explore two of such examples: first we study a flavor specific Two-Higgs Doublet Model (2HDM)~\cite{Botella_2020,Pich_2009,Celis_2013}, where the chiral enhancement is obtained only at the two-loop level by the s.c. Barr-Zee mechanism. Secondly, we study two scalar LeptoQuarks, namely $S_1$ and $R_2$ \cite{Dor_ner_2016}, that can generate the desired operators already at tree level. In both cases we compare the RGE improved SMEFT results to the explicit (unresummed) model calculations and discuss the validity of the approximations taken in each approach.

The paper is organized as follows. In Section~\ref{sec:basis} we present the dimension six operators relevant for the EFT analysis and describe their interplay due to the RGE. Section~\ref{sec:matchingbelowEW} is devoted to matching the SMEFT operators to observables below the EW scale. The phenomenological aspects of the latter are discussed in Section~\ref{sec:pheno:intro} for the three charged lepton flavors. In Section~\ref{sec:Model} we compare the SMEFT results to explicit calculations within a flavor specific 2HDM, while in Section~\ref{sec:LeptoQuarks} we do the same for the scalar LeptoQuarks case. We then summarize our conclusions in  Section \ref{sec:conclusions}. Finally, Appendix~\ref{app:oneloop} gives details on the RG equations for the chosen operators, Appendix~\ref{app:oneloopcalc} gives details of our calculation of the one-loop SMEFT matching when integrating out the top quark, while Appendix \ref{app:leadlog} describes the leading logarithm expansion of two-loop diagrams in the 2HDM.

\section{SMEFT Operator Basis}
\label{sec:basis}
We start by building the SMEFT at scales above the EW symmetry breaking (EWSB) scale, $v = 246$\,GeV by extending the SM Lagrangian with a series of operators of increasing canonical dimension ($d$): $\mathcal L = \mathcal L^{\rm SM} + \sum_d\mathcal L^{(d)}$. The most relevant dimension-six Lagrangian can be written as
\beq
{\cal L}^{(6)} = \sum_i \hat\C_i \op_i + {\rm h. c.} \,,\qquad{\rm with}\qquad \hat\C_i=\frac{\C_i}{\Lambda^2}\,,
\eeq
where the index $i$ runs over all the SM gauge invariant operators in a given basis. In the following we work with the s.c. ``Warsaw'' basis of operators~\cite{Grzadkowski_2010}. For simplicity, since we are concerned with phenomenology of CP-even observables, we limit our discussion to CP conserving NP dynamics. In practice, i.e. for the operators we consider, this means taking all $\hat C_i$ as real. 

The set of operators relevant for the analysis of charged lepton anomalous magnetic moments and leptonic Higgs decays is actually quite small, including the three operators
\begin{subequations}
\begin{align}
{\cal O}_{1,pr} &= \lp \varphi^\dagger\varphi \rp \lp \bar \ell_p e_r \varphi \rp\,, \label{eq:operators:1}\\
{\cal O}_{2,pr} &=  \lp \bar \ell_p \sigma^{\mu\nu} e_r  \rp \tau^a \varphi W_{\mu\nu}^a\,,\label{eq:operators:2}\\
{\cal O}_{3,pr} &=  \lp \bar \ell_p \sigma^{\mu\nu} e_r  \rp \varphi B_{\mu\nu}\,.\label{eq:operators:3}
\end{align}
\end{subequations}
Here $\varphi$ is the Higgs doublet, $\ell$ is the left-handed lepton doublet, $e$ is the right-handed lepton singlet and $p,r$ are the lepton generation indices. $W$ and $B$ are the $SU(2)_L$ and $U(1)_Y$ gauge field strength tensors respectively, while $\tau^a$ are the $SU(2)_L$ generators.

As the relevant scales for the evaluation of $g-2$ and Higgs decays are below the EW and especially NP scales (where we assume $v \ll \Lambda$), the RG evolution and matching of operators from the NP scale to low energies needs to be taken into account. The complete system of one-loop SMEFT RG equations is presented in Refs.~\cite{Jenkins_2013,Jenkins_2014,Alonso_2014}. The important observation is that, setting all the fermion Yukawa couplings except for the top quark to zero, operators in Eqs.~(\ref{eq:operators:1} - \ref{eq:operators:3}) can form a closed set under RGE provided we include two additional four-fermion operators
\beq\label{eq:operators:4and5}
\op_{4,prst} = \lp \bar\ell_p^j e_r \rp\epsilon_{jk} \lp\bar q_s^k u_t \rp\,,\quad\op_{5,prst} = \lp \bar\ell_p^j \sigma_{\mu\nu}e_r \rp\epsilon_{jk} \lp\bar q_s^k\sigma^{\mu\nu} u_t \rp\,,
\eeq
where $q$ and $u$ are the quark LH doublet and up-quark RH singlet respectively, with generation indices $r, t$. The indices $j,k$ are $SU(2)_L$ gauge indices, contracted by the antisymmetric Levi-Civita tensor $\epsilon_{jk}$. In the following, we will consider only operators including the top-quark and conserving lepton flavor, thus we suppress the $s=t$ indices for quarks and $p=r=e,\mu,\tau$ indices for leptons.  Finally, to form a closed set of RG equations, we need to take into account the RG evolution of gauge and Yukawa couplings. The complete set of RG equations we use is collected in Appendix \ref{app:oneloop}.

In the approximations described before, the gauge couplings, the Higgs mass and the top quark Yukawa evolve according to the SM equations, which we take from Refs.~\cite{Mihaila_2012} and \cite{Chetyrkin_2012}, respectively. On the other hand, the Higgs quartic coupling $\lambda$ and the muon Yukawa $y_\mu$ receive one-loop contributions from the operators ${\cal O}_1$ and ${\cal O}_4$, described in Eqs.~\eqref{eq:oneloop:muonyuk} and~\eqref{eq:oneloop:quartic}. The former contains corrections to $y_\mu$ which, differently from the SM, are not proportional to the $y_\mu$ Yukawa itself. This is also reflected in the one-loop threshold corrections at the weak scale proportional to $v^2\C_{i,p}$. The corrections to $\lambda$ are instead also proportional to $y_\mu$ and we keep them for completeness, but are otherwise negligible.

The RG evolution of these operators can hint at the low-energy signatures, which will be calculated in detail in the next Section. As an illustration of the interplay between them, we set the initial conditions at the high scale $\Lambda = 10^8 ~ {\rm GeV}$ and solve the RG evolution down to the weak scale $v$. In each case, we set $\C_{i,\ell}(\Lambda) = 1$ while $\C_{j,\ell}(\Lambda) = 0$ for $j\neq i$. In other words, we consider the cases where the UV complete model generates one operator for a specific flavor $\ell$ at the leading EW order, while all the others are EW loop-induced. The results are shown in Fig.~\ref{fig:running}. 

Note that $\C_{1,\ell}$ does not appear in the RG equations for $\C_{2,3,4,5,\ell}$ (see Appendix \ref{app:oneloop}); this is reflected in the first plot in Fig.~\ref{fig:running}, where the one-loop induced coefficients are all negligible. Thus, UV models that only generate ${\cal O}_{1,\ell}$ can be probed exclusively via Higgs decays. 

On the other hand, $\C_{2,\ell}$ and $\C_{3,\ell}$ can induce a sizeble $\C_{1,\ell}$ through their RG evolution. It follows that bounds on leptonic Higgs decays become relevant for these models, together with bounds on the lepton anomalous magnetic moments, which ${\cal O}_{2,\ell}$ and ${\cal O}_{3,\ell}$ generate at the leading order when matched to the low energy EFT below the EW scale (see next Section for details).

Finally, the two four-fermion operators ${\cal O}_{4,\ell}$ and ${\cal O}_{5,\ell}$ are able to induce $\C_{1,\ell}$, $\C_{2,\ell}$ and $\C_{3,\ell}$, however in very different proportions. This behaviour allows one-loop matching contributions (of ${\cal O}_{5,\ell}$) to the anomalous magnetic moment at the low scale to dominate over the tree-level ones (from $\C_{2,\ell}$ and $\C_{3,\ell}$), while also generating small deviations in Higgs decays (due to $\C_{1,\ell}$). Thus, the case of ${\cal O}_{5,\mu}$ looks particularly promising, since it can address the $a_\mu$ problem at relatively low NP scales, which in turn can be probed by near future experiments measuring $h\to\mu\mu$ decays.
\begin{figure}[t]
	\centering
	\includegraphics[width=0.47\linewidth]{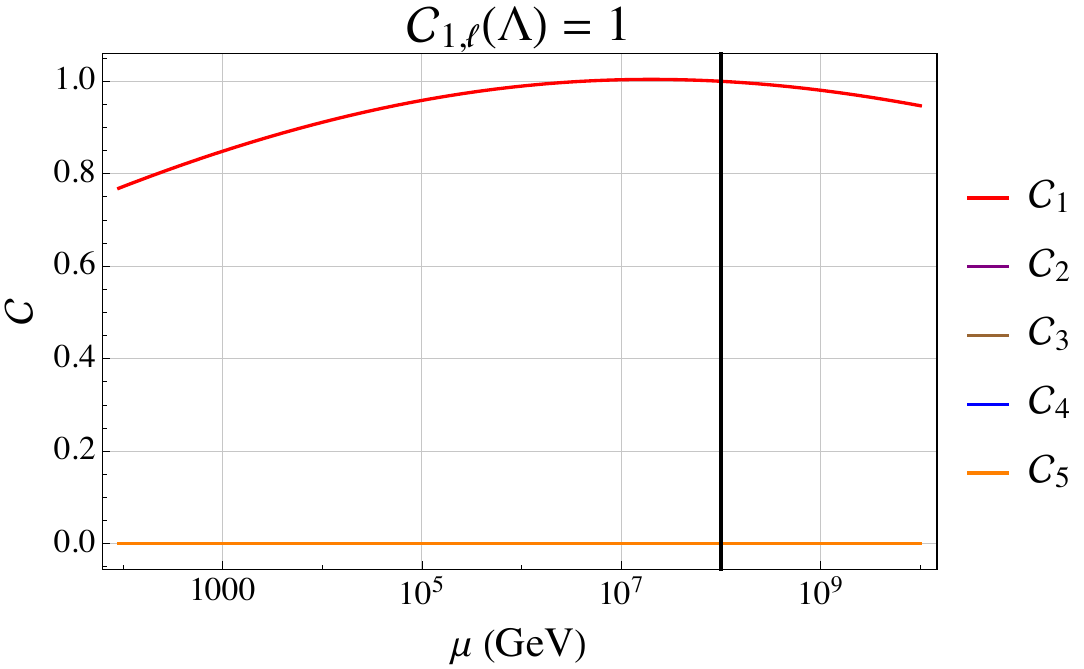} 
	\includegraphics[width=0.47\linewidth]{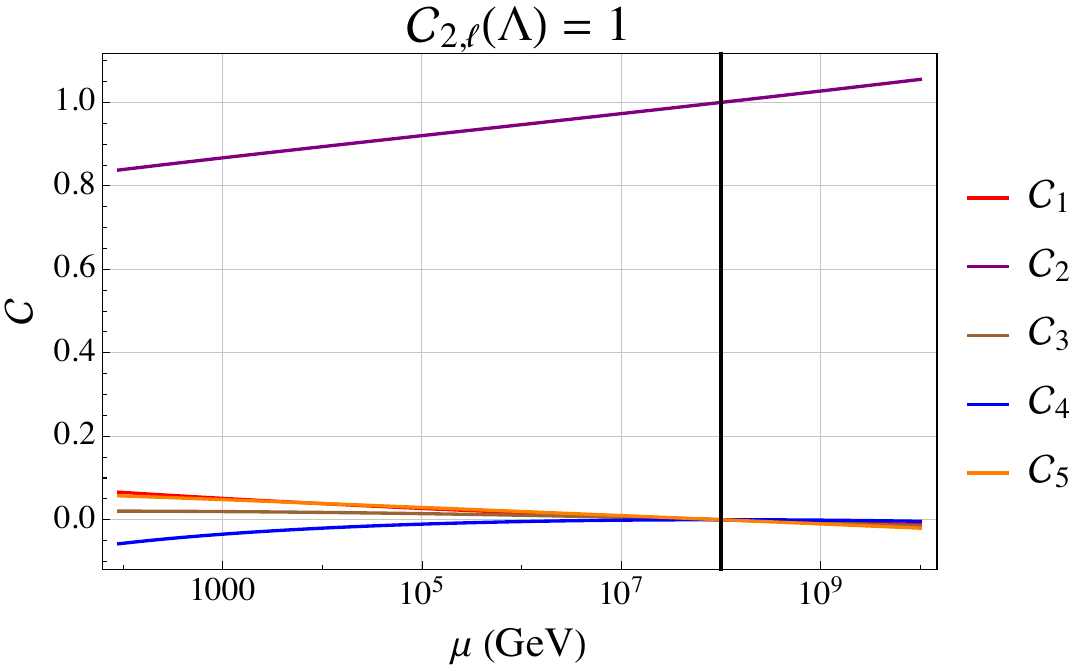}
	\includegraphics[width=0.47\linewidth]{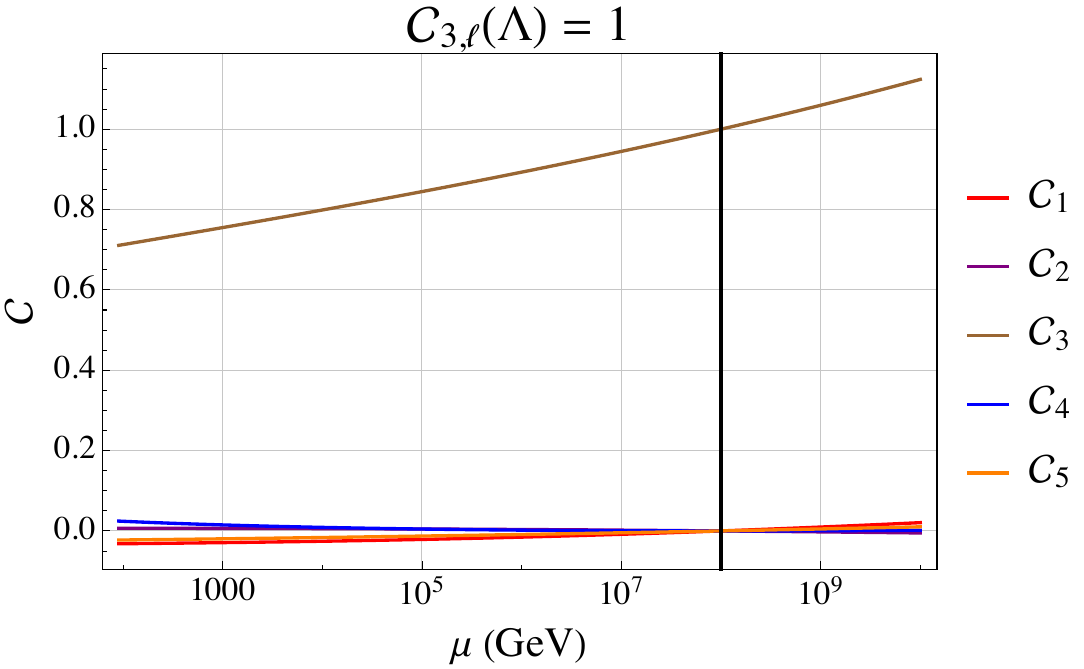} 
	\includegraphics[width=0.47\linewidth]{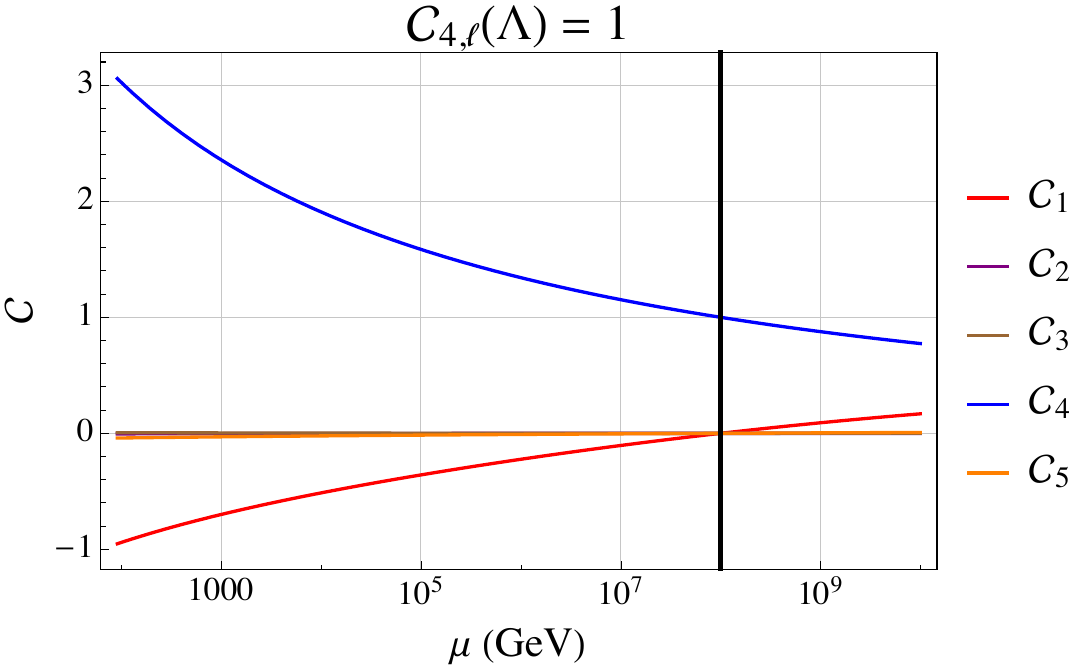}
	\includegraphics[width=0.47\linewidth]{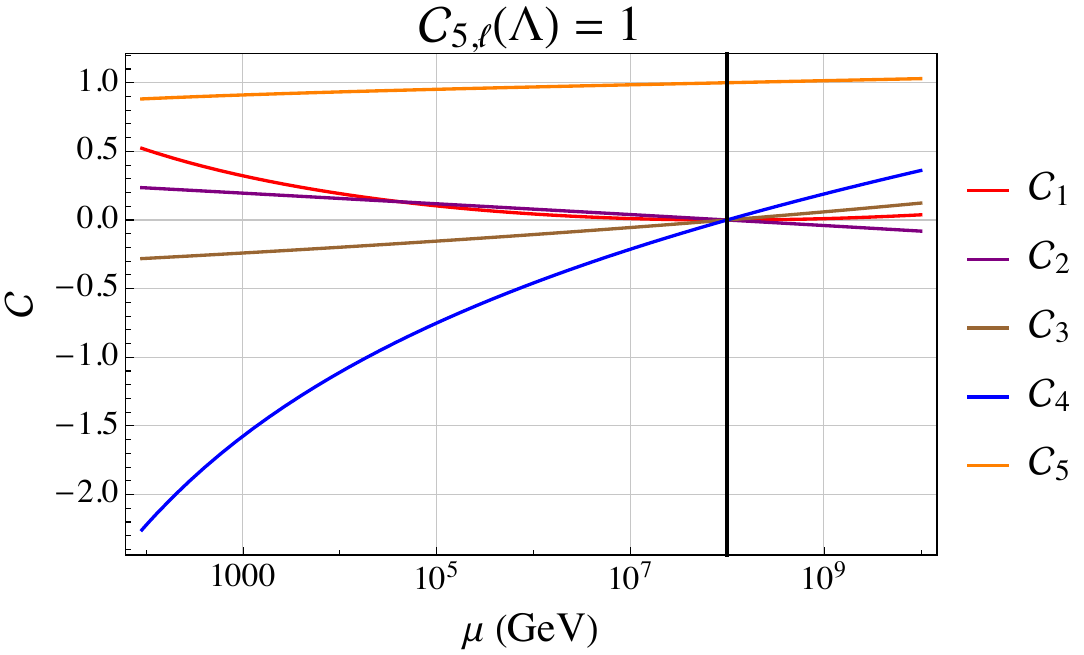}
	\caption{Running for the five different initial conditions $\C_{i,\ell}(\Lambda) = 1$. The color coding for each Wilson coefficient is the same in each plot and shown in the legend on the plot side. The vertical black line is at $\Lambda = 10^8$ GeV.}\label{fig:running}
\end{figure} 

\section{Operator Matching to Observables Below EWSB }\label{sec:matchingbelowEW}
Below the scale of EWSB we can integrate out the heavy degrees of freedom, namely the top quark $t$ and the weak gauge bosons $W$ and $Z$. The Higgs doublet acquires the form 
\beq
\varphi^T = \frac{1}{\sqrt{2}} \lp 0 , h + v \rp\,.
\eeq
We keep the dynamical field $h$ in the discussion of Higgs decay only, while we can integrate it out when treating the lepton anomalous magnetic moment.

The operators entering the low energy effective Lagrangian ($\mathcal L \ni \sum_i \mathcal Q_i$) relevant for our phenomenological analysis (assuming CP invariance) below the EWSB scale only respect EM (and QCD) gauge invariance 
\beq\label{eq:basis:belowEW}
{\mathcal Q}_{1,\ell} =   { c_{1,\ell}}{} h \lp\bar l_\ell  l_\ell\rp\,, \qquad {\mathcal Q}_{2,\ell} = e \frac{ {  c}_{2,\ell}}{m_\ell} \lp\bar l_\ell \sigma_{\mu\nu}  l_\ell \rp F^{\mu\nu}\,,
\eeq
where now $l_\ell$ denotes the $\ell$-flavored charged lepton, $F^{\mu\nu}$ stands for the photon field strength and $e = \sqrt {4\pi\alpha}$ is the QED gauge coupling. 
The first operator induces Higgs decay to two leptons, while the second generates the lepton anomalous dipole moment.

When matching the low energy operators to the SMEFT basis, it is important to consider and reduce the matching scale ($\mu_w$) dependence of the physical results, by matching to sufficiently high loop order. To parametrize the relative importance of matching scale and loop corrections, we define the ratio
\beq
R_{i,j}(\mu_w) = \frac{c_{i,\ell}(\mu_w)|_j}{c_{i,\ell}(m_t)|_L}\,, \quad i=1,2\,,
\eeq 
where the label $j = T,~L$ refers to matching to SMEFT at tree level only (T) or including 1-loop terms (L). Fixing the SMEFT NP scale to $\Lambda = 10^4$ GeV and varying the matching scale $\mu_W$ between the $W$ mass and the EW vev $v$,  the ratios for the two low energy operators are plotted in Fig.~\ref{fig:Ratios}.  We discuss the details of this matching and its implications for phenomenology associated with each operator separately below.

\subsection{Higgs decay to two leptons}
\label{subsec:higgsdecay}
In the SM the Higgs boson interacts with leptons through the Yukawa terms in the Lagrangian
\beq
{\cal L} \supset y_{\ell}^{SM} \bar \ell_\ell \varphi e_\ell + {\rm h. c.} = y_{\ell}^{SM} \lp\frac{h + v}{\sqrt{2}}\rp \bar l_\ell  l_\ell\,,
\eeq
where we wrote the explicit form of the neutral component of the Higgs doublet below the EWSB scale. The Yukawa couplings in the SM are $y_{\ell}^{SM} = m_\ell\sqrt{2}/v$. 

The effect of the renormalizable operator ${\mathcal Q}_{1,\ell}$ can be included in a straightforward way as
\beq\label{eq:Hdecay:lowEFT}
{\cal L} \supset \lp \frac{m_\ell}{v}  + \delta{c_{1,\ell}}{} \rp  {h}{} \bar l_\ell l_\ell\,.
\eeq
At the matching scale ($\mu_{w}$) below EWSB, (real) $\C_{1,\ell}$ matches at tree level into $c_{1,\ell}$ as
\beq
\delta c_{1,\ell}(\mu_w)|_{tree} = {\hat \C_{1,\ell}(\mu_w)}{} \frac{ v^2}{\sqrt{2}}\,.
\eeq

\begin{figure}[t]
	\centering
	\includegraphics[width=1\linewidth]{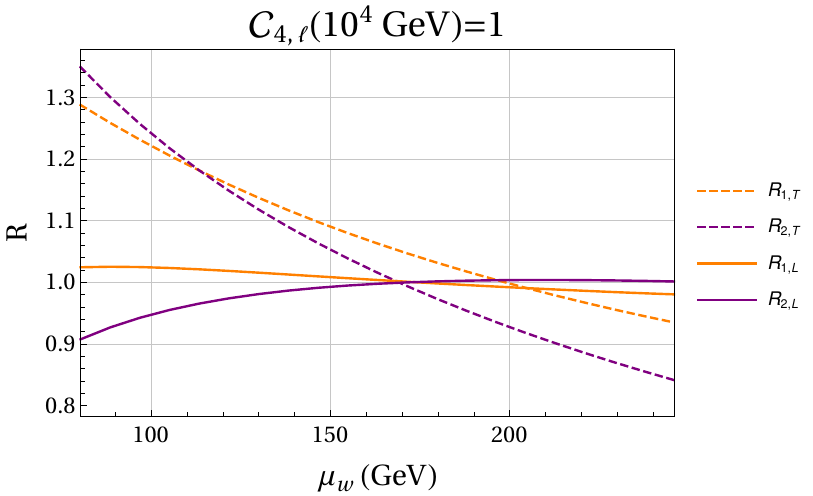} 
	\caption{Ratios of matching conditions as function of the matching scale below EWSB $\mu_W$ for fixed $\Lambda = 10^4$ GeV and $\C_{4,\ell}(\Lambda) = 1$ as the initial condition for the RGE.}\label{fig:Ratios}
\end{figure}

Further threshold corrections from top quark loops, induced by ${\cal O}_{4,\ell}$, appear at next to leading order. The details of the calculation are given in Appendix \ref{app:oneloopcalc:O4O2}. 
The resulting matching correction is of the form
\beq\label{eq:deltay:matching}
\delta c_{1,\ell}(\mu_w)|_{loop} = \frac{N_c m_t^3}{8 \pi^2 v} \hat\C_{4,\ell}(\mu_w) \left[ \frac{1}{3} +  ~ {\rm ln}\lp \frac{m_t}{\mu_w}   \rp \right]   \,.
\eeq
The importance of including the one-loop contribution can be understood from comparing $R_{1,T/L}(\mu_w)$  (orange lines) plotted in Fig.~\ref{fig:Ratios}. We observe that the tree-level matched result has a sizable $\mu_w$ dependence: starting with a fixed value of $C_{4,\ell}$ at the high scale of $\Lambda=10^4$\,GeV,  $c_{1\ell}$ can change by almost a factor of 1.5 when varying $\mu_w \in [m_W,v]$ around the top mass\,. Including the one-loop correction, the scale dependence is greatly reduced to around $5\%$\,.

Collider experiments measure the signal strenght of leptonic Higgs decays in various Higgs production modes. These can in turn be related to the so-called $\kappa$ formalism, where the relevant quantity is defined as
\beq\label{eq:Hdecay:ratio}
\kappa_\ell^2 = \frac{\Gamma(h\to\ell^+\ell^-)}{\Gamma_{SM}(h\to\ell^+\ell^-)}\,,
\eeq
where we can now identify
\beq\label{eq:signalstrenght:firstorder}
\kappa_\ell = \lp 1 + \frac{ \delta c_{1,\ell} v}{m_{\ell}} \rp \,,
\eeq
where we evaluate $\delta c_{1,\ell}$ at the Higgs mass ($\mu_w = m_h$).

\subsection{Anomalous magnetic moment}
\label{subsec:anomalousmoment}
Measurements of $a_\ell$ are generally performed at the lepton mass scale, at which also the Higgs boson can be integrated out safely.\footnote{The case of $a_\tau$ is slightly different but still consistent with this assumption, see the discussion in Sec.~\ref{subsec:pheno:tau}. } At tree level the operators ${\cal O}_{2,\ell}$ and ${\cal O}_{3,\ell}$ contribute to the dipole operator ${\mathcal Q}_{2,\ell}$ by projecting out the photon field, while the Higgs acquires a vev. The matching condition (for real $\mathcal C_{2,3}$) then reads
\beq
\delta c_{2,\ell}(\mu_w)|_{tree} = \frac{v \, m_\ell }{\sqrt{2}e} \lp c_w \hat\C_{3,\ell}(\mu_w) - s_w \hat\C_{2,\ell}(\mu_w) \rp\,,
\eeq
where $c_w$ and $s_w$ are the cosine and sine of the weak mixing angle respectively.

At the next to leading order, the threshold corrections come from two different diagrams. On one hand, we have ${\cal O}_{5,\ell}$ matching into ${\mathcal Q}_{2,\ell}$, again through a top quark loop. This  numerically dominant contribution is evaluated explicitly in Appendix~\ref{app:oneloopcalc:O5O1}. On the other hand, additional corrections arise when we integrate out the $W$, $Z$ and the Higgs. 
The complete matching condition with the low energy operator was recently computed in Ref.~\cite{Dekens_2019} and includes EW renormalization of ${\cal C}_{2,\ell}$ and ${\cal C}_{3,\ell}$ as well as of the EW vev, the lepton mass and $e$.
The end result is
\beq\label{eq:matching:Cprime}
\delta c_{2,\ell}(\mu_w)|_{1-loop} = - \frac{N_c Q_t \hat\C_{5,\ell}(\mu_w) m_t m_\ell}{2 \pi^2 } {\rm ln}\lp\frac{\mu_w}{m_t} \rp + {\cal O}(\alpha)\,,
\eeq
where $Q_t=2/3$ the top-quark electric charge. 
To abbreviate the expression, the $ {\cal O}(\alpha)$ term stands for all the one-loop contributions from ${\cal O}_{2,\ell}$ and ${\cal O}_{3,\ell}$, which we take into account in our analysis but are numerically sub-dominant (they include both finite threshold effects as well as logarithmic scale dependent terms). Again we highlight the importance of including the one-loop matching contributions in Fig.~\ref{fig:Ratios}, where we plot $R_{2,T/L}(\mu_w)$ (purple lines). We observe that even more than in the case of $c_{1\ell}$ the tree-level matched result has a pronounced $\mu_w$ dependence: starting with a fixed value of $C_{4,\ell}$ at the high scale of $\Lambda=10^4$\,GeV,  $c_{2\ell}$ can change by almost a factor of two when varying $\mu_w \in [m_W,v]$ around the top mass\,. Including the one-loop correction, the scale dependence is greatly reduced to below $10\%$\,.
 
The subsequent running from the weak scale to the lepton mass scale is driven by QED interactions only. The are expected to induce a shift in $c_{2\ell}$ of at most a few percent, and so in light of the residual matching scale variance can be safely neglected. In our phenomenological analysis we thus take $\delta  c_{2,\ell}(m_t) \simeq \delta  c_{2,\ell}(m_\ell)$, such that the matching is numerically dominated by the tree-level contributions. The shift of the lepton anomalous magnetic moment will then simply be 
\beq
\delta a_\ell = {4 \delta c_{2,\ell}(m_\ell)}{}\,.
\eeq

\section{Phenomenology}
\label{sec:pheno:intro}
In the section we explore the phenomenological implications of NP parametrized by the operators in Eq. \eqref{eq:basis:belowEW}. Given the initial conditions for the Wilson coefficients $\C_{i,\ell}$ at a high scale $\Lambda$, we use the set of RG equations in Appendix \ref{app:oneloop} to evolve these coefficients down to the weak scale $\mu_w \sim m_t$. We then use the matching conditions in Eq. \eqref{eq:matching:Cprime} and \eqref{eq:deltay:matching} to get the low energy coefficients, which can be directly used to compare with experimental results. Note that with the assumption of massless leptons (above EWSB) and no running below the weak scale, the RG evolution itself is lepton flavor-independent given the same initial conditions. 

The effect on low-energy observables depends on the initial conditions of the high-energy Wilson coefficients at the scale $\Lambda$. In this section we present the results corresponding to the option where all coefficients but one are zero at $\Lambda$, i.e. the UV complete model generates one operator, while all the others are induced by the EW RG evolution. Other combinations are also possible and we explore this possibility using explicit UV completions in Secs.~\ref{sec:Model} and~\ref{sec:LeptoQuarks}.  

\subsection{Electron}
\label{subsec:pheno:electron}
\begin{figure}[t]
	\centering
	\includegraphics[width=1\linewidth]{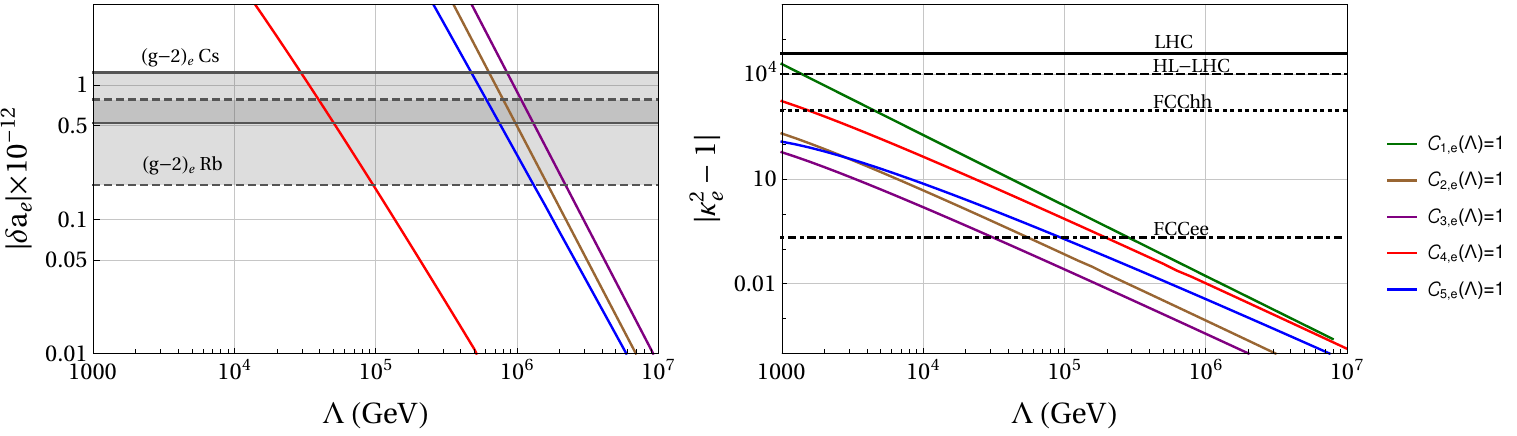} 
	\caption{\textbf{Left}: Induced shift in the electron anomalous magnetic moment from $\delta c_{2,e}$ as function of the NP scale $\Lambda$. The horizontal gray bands show the $(g-2)_e$ favored regions according to Cs (solid) and Rb (dashed) measurements. \textbf{Right}: Induced shift in Higgs to two electrons decay from $\delta c_{1,e}$ as function of the NP scale $\Lambda$. The horizontal solid line shows present bounds from measurements at LHC, while dashed, dotted and dotdashed lines show projections from HL-LHC, FCChh and FCCee respectively.  }\label{fig:electron:all}
\end{figure}

The electron anomalous magnetic moment is one of the most precisely measured quantities in particle physics. Taking into account recent (diverging) measurements of the fine structure constant in Cs~\cite{Parker:2018vye} and Rb~\cite{Morel:2020dww} atoms, the SM predicted and measured value of $a_e$ differ by
\beq\label{eq:electronanomalousmoment}
\begin{split}
\delta a_e^{Cs} &= a_e^{\rm exp } - a_e^{\rm SM(Cs)} = \lp-0.88\pm0.36\rp\times10^{-12}\,, \\
\delta a_e^{Rb} &= a_e^{\rm exp } - a_e^{\rm SM(Rb)} = \lp0.48\pm0.30\rp\times10^{-12}\,, \\
\end{split}
\eeq
and thus currently exhibit a $\sim2.4~\sigma$ and $\sim1.6~\sigma$ tension, according to the Cs and Rb measurement of $\alpha$, respectively. In Fig.~\ref{fig:electron:all} we compare these ranges to the predicted shift in $a_e$ due to different SMEFT operators.

A measurement of Higgs decay to two electrons within the SM is outside the reach of present experiments. This is a consequence both of low rates of Higgs production at $pp$ colliders but especially the smallness of the Higgs coupling to electrons in the SM, $y^{\rm SM}_e = m_e v/\sqrt{2}\sim3\times10^{-6}$.
While ATLAS and CMS already provided upper bounds on $\kappa_e$ \cite{Khachatryan:2014aep,Aad_2020} of the order of $10^2$, neither the high luminosity (HL) LHC upgrade nor a higher energy $pp$ collider like the FCC-hh would be able to improve these considerably. Future ``Higgs factories'' might thus be required to achieve a significant progress in this decay channel. As reported in Ref.~\cite{blondel2020circular}, for example, FCC-ee (high energy circular electron-positron collider) would be able to probe $\kappa_e$ at the $10\%$ level. These present and projected bounds are shown in the right plot of Fig.~\ref{fig:electron:all}.

As expected, both LHC and HL-LHC are not able to exclude NP scales above a TeV. Even projections from FCC-hh are only able to exclude NP below few TeV in $h\to ee$ induced at tree-level by $\C_{1,e}$. Furthermore, $\C_{1,e}$ cannot be probed by measurements of $a_e$, thus making such NP effects very challenging to probe. Conversely, out of all considered scenarios addressing the present $a_e$ discrepancy, only the ${\cal C}_{4,e}(\Lambda)=1$, $\Lambda = $\,few\,$\times 10^{4}$\,GeV  case has the potential to be tested through $h\to e^+e^-$ decay measurements at the FCC-ee. 

\subsection{Muon}
\label{subsec:pheno:muon}
\begin{figure}[t]
	\centering
	\includegraphics[width=1\linewidth]{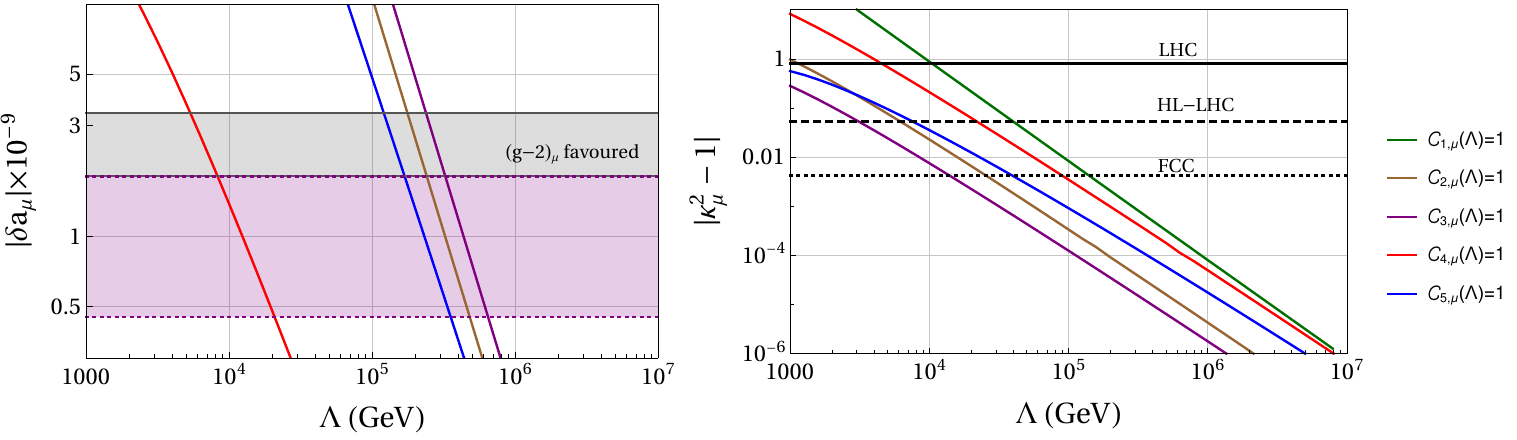} 
	\caption{\textbf{Left}: Induced shift in the muon anomalous magnetic moment from $\delta c_{2,\mu}$ as function of the NP scale $\Lambda$. The horizontal gray and purple bands show the $(g-2)_\mu$ favored region assuming $\delta a_\mu$ as in Eq. \eqref{eq:muonanomalousmoment:PDG} or in Eq. \eqref{eq:muonanomalousmoment:Lattice} respectively. \textbf{Right}: Induced shift in Higgs to two muons decay from $\delta c_{1,\mu}$ as function of the NP scale $\Lambda$. The horizontal solid line shows present bounds from measurements at LHC, while dashed and dotted lines show projections from HL-LHC and FCC respectively.}\label{fig:muon:all}
\end{figure}

The muon anomalous magnetic moment exhibits a long standing tension between its measured and predicted value. Currently it is estimated at $\sim 3.3 \sigma$ or numerically~\cite{PhysRevD.98.030001}
\beq\label{eq:muonanomalousmoment:PDG}
\delta a_\mu = a_\mu^{\rm exp} - a_\mu^{\rm SM} = \lp261\pm79\rp\times10^{-11}\,,
\eeq
where the error represents the combined theoretical and experimental uncertainties.  Note that the recent BMWc Lattice result~\cite{Borsanyi:2020mff} updates the value of the leading order hadron vacuum polarization (LO-HVP) contribution. When combined with the other terms described in Ref.~\cite{PhysRevD.98.030001}, we get
\beq\label{eq:muonanomalousmoment:Lattice}
\delta a_\mu = \lp113\pm68\rp\times10^{-11}\,,
\eeq
which corresponds to a $\sim1.7\sigma$ deviation. 
Left plot in Fig.~\ref{fig:muon:all} shows the shift in $a_\mu$ induced by the dipole operator ${\cal Q}_{2,\mu}$ compared to the discrepancy in Eqs.~\eqref{eq:muonanomalousmoment:PDG} (gray region) and~\eqref{eq:muonanomalousmoment:Lattice} (purple region), when different initial conditions at the high-energy scale $\Lambda$ are taken.
Again notice that models that only generate ${\cal O}_{4,\mu}$ in the UV can account for the discrepancy in Eq.~\eqref{eq:muonanomalousmoment:PDG} only at a relatively low scale interval, $3 \lesssim\Lambda \lesssim7$ TeV, while other scenarios point to much larger NP scales above 100~TeV. One would instead require scales of $10 \lesssim\Lambda \lesssim20$ TeV in order for ${\cal O}_{4,\mu}$ to account for the discrepancy in Eq.~\eqref{eq:muonanomalousmoment:Lattice}.

Recently, di-muon Higgs decay was directly observed for the first time at LHC. ATLAS and CMS reported the values \cite{atlascollaboration2020search}
\beq
{\rm ATLAS}: \kappa_\mu^2=1.2\pm 0.6\,, \qquad {\rm CMS}: \kappa_\mu^2=1.19 \pm 0.55\,,
\eeq
for the signal strength. The precision of these measurements is expected to be improved significantly at the HL-LHC~\cite{cepeda2019higgs} and especially FCC~\cite{Mangano:2651294}. Converting these projections to our notation, 
the reported values imply
\beq
{\rm HL-LHC}: |\kappa_\mu^2-1| < 0.053\,, \qquad {\rm FCC}: |\kappa_\mu^2-1| < 4.2\times10^{-3}\,.
\eeq
The results are shown in the right plot in Fig.~\ref{fig:muon:all}.  We observe that compared to the NP sensitivity of $a_\mu$, the current LHC Higgs decay measurements are not yet competitive in any of the considered scenarios. However, already the HL-LHC upgrade could potentially probe the current $a_\mu$ discrepancy in the ${\cal C}_{4,\mu}(\Lambda)=1$ scenario. Sensitivity to other scenarios would unfortunately be marginal even at the FCC. Better direct sensitivity could possibly be achieved at a dedicated muon collider~\cite{capdevilla2021nolose,buttazzo2020probing,yin2021muon}.


\subsection{Tauon}
\label{subsec:pheno:tau}
\begin{figure}[t]
	\centering
	\includegraphics[width=1\linewidth]{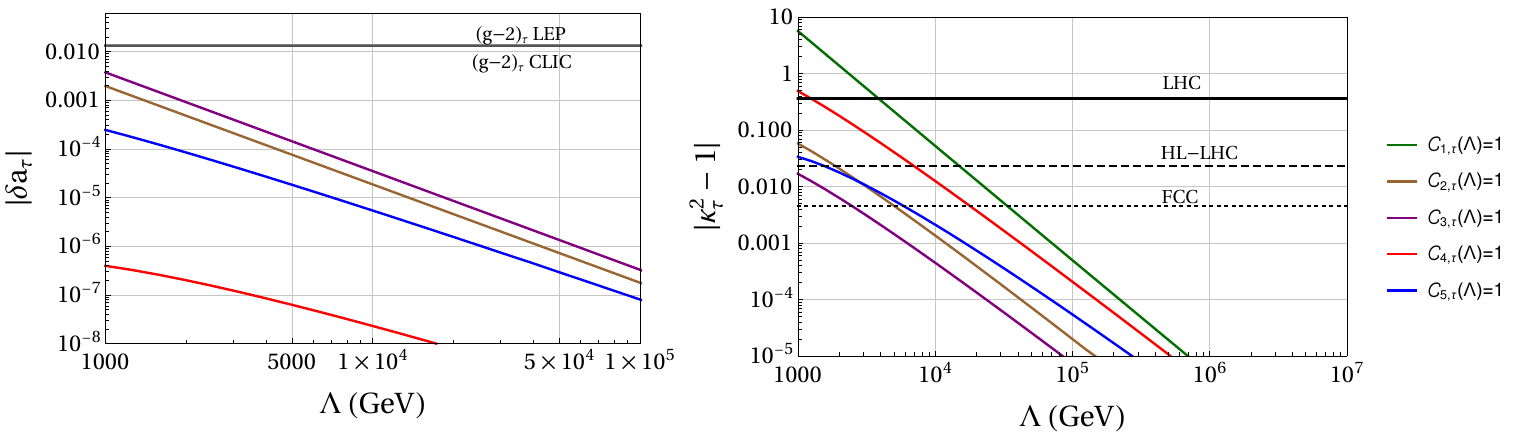} 
	\caption{\textbf{Left}: Induced shift in the tau anomalous magnetic moment from $\delta c_{2,\tau}$ as function of the NP scale $\Lambda$. The horizontal lines show the upper bounds from LEP and CLIC, see Eq. \eqref{eq:bounds:atau}. \textbf{Right}: Induced shift in Higgs to two taus decay from $\delta c_{1,\tau}$ as function of the NP scale $\Lambda$. The horizontal solid line shows present bounds from measurements at LHC, while dashed and dotted lines show projections from HL-LHC and FCC respectively.}\label{fig:tau:all}
\end{figure}

The case for the anomalous magnetic moment of the tau lepton, $a_\tau$, is quite different than the previous two. The very short tau lifetime does not allow for precise measurement of its spin precession in a magnetic field, and $a_\tau$ needs to be extracted from the cross section of $\tau$-pair production in high energy processes. These typically involve (virtual) photons with non-negligible $q^2$ and in practice probe the EM form factor of the tau away from $q^2=0$ limit where $a_\ell$ is conventionally defined. The SM predicted value (at $q^2=0$)  is~\cite{Keshavarzi_2020}
\beq
a_\tau^{\rm SM} = (117717.1 \pm 3.9) \times 10^{-8}\,,
\eeq
while experimental limits from LEP~\cite{Carrington_2004} and projections from CLIC~\cite{Billur_2014}\footnote{The bounds refer to 200 fb$^{-1}$ luminosity, $\sqrt{s} = 1.5$ TeV and $\delta_{sys} = 0.1$\,.} are:
\beq\label{eq:bounds:atau}
{\rm LEP}: -0.052 < a_\tau^{\rm exp} < 0.013\,, \qquad {\rm CLIC}: -0.0131 < a_\tau^{\rm exp} < 0.0091\,.
\eeq
First, we observe that the experimental sensitivity is not expected to reach the order of magnitude of the SM predictions (which have non-negligible $q^2$ dependence) in the foreseeable future. In addition, as discussed in Sec.~\ref{subsec:anomalousmoment}, contributions due to heavy NP experience negligible running below the EWSB scale. Consequently, the experimental bounds on $\delta a_\tau$ at any small enough $q^2$ will apply equally and can be used to constrain the relevant UV scenarios. The results are shown in the left plot in Fig.~\ref{fig:tau:all}. Unfortunately, none of the considered scenarios with $\Lambda$ above the TeV is expected to yield observable signals at present or planned experiments. Any prospective signal in such measurements would thus point to the presence of light new degrees of freedom.


On the other hand, $h\to\tau\tau$ has been measured with good precision at LHC. The signal strength reported by ATLAS~\cite{Aaboud:2018pen} and CMS~\cite{Sirunyan_2018} are
\beq
{\rm ATLAS}: \kappa_\tau^2=1.09\pm 0.29\,, \qquad {\rm CMS}: \kappa_\tau^2=1.09 \pm 0.27\,.
\eeq
Similar to the muon case, projected bounds are considered from HL-LHC~\cite{cepeda2019higgs} and FCC~\cite{Mangano:2651294} and can be translated to our notation as
\beq
{\rm HL-LHC}: |\kappa_\tau^2-1| < 0.023\,, \qquad {\rm FCC}: |\kappa_\tau^2-1| < 4.5\times10^{-3}\,.
\eeq
The results are shown in the right plot in Fig.~\ref{fig:tau:all}. We observe that the $\mathcal C_{1,\tau}(\Lambda)=1$ scenario could be probed up to $\Lambda \sim 10$\,TeV already at the HL-LHC, while the other scenarios would require the FCC to be probed experimentally to the same degree, with $\mathcal C_{2,\tau}$ case being most elusive.

\section{Flavor specific 2HDM}\label{sec:Model}
Based on the results of the EFT analysis, it is interesting to consider UV completions that generate one of the operators, e.g. ${\cal O}_{4,\ell}$\,, above the EW scale already at the tree level, while others are EW loop induced. In this section we consider a minimal Two Higgs-Doublet Model (2HDM) extension of the SM\footnote{For reviews of the topic, see e.g. Refs.~\cite{Branco_2012,BHATTACHARYYA_2016,Celis_2013}.}, where we restrict the interactions of the new heavy scalars to only one charged lepton flavor (the muon), as well as to one up-like quark flavor (the top).

\subsection{Model description}
\label{sec:2HDM}
We introduce two $SU(2)_L$ doublets, $\varphi_1$ and $\varphi_2$, with hypercharge +1 and vacuum expectation values (vevs) $v_1$ and $v_2$, respectively. This sector can be rotated via a global $SU(2)$ transformation in order to make $\varphi_1$ to act as the SM Higgs~\cite{Pich_2009}. We have 
\beq
\begin{pmatrix} \Phi_1 \\ \Phi_2 \end{pmatrix} = \begin{pmatrix} \cos\beta & \sin\beta \\  -\sin\beta & \cos\beta \end{pmatrix} \begin{pmatrix} \varphi_1 \\ \varphi_2 \end{pmatrix}\,, \qquad \tan\beta=\frac{v_2}{v_1}\,.
\eeq
In this basis, named the Higgs basis, $\Phi_2$ has no vev while $\Phi_1$ acquires a vev $v = \sqrt{v_1^2 + v_2^2}$. To simplify our discussion, we take from the start $v_2=0$ and $v_1=v$; this corresponds to the case $\sin\beta=0$ and $\cos\beta=1$.

In the Higgs basis we can decompose the doublets as
\beq
\Phi_1 = \begin{pmatrix} G^+ \\ \frac{v + \rho_1 + i G_0}{\sqrt{2}} \end{pmatrix}\,, \qquad \Phi_2 = \begin{pmatrix} H^+ \\ \frac{\rho_2 + i \eta}{\sqrt{2}} \end{pmatrix}\,.
\eeq
where $\rho_{1,2}$ and $\eta$ are the neutral scalar and pseudoscalar components respectively and $H^+$ is the charged degree of freedom. This global transformation has the advantage of clearly isolating the Goldstone bosons $G^\pm$ and $G^0$ in the decomposition, which will be eaten to give mass to $W^\pm$ and $Z^0$.

The scalar fields $\rho_{1,2}$ can be additionally related with the physical Higgs field and a heavy neutral scalar by an orthogonal transformation. Here we assume that the scalar potential is CP conserving, thus the pseudoscalar components do not mix with the scalar one. The CP-even mass eigenstates can be then obtained by a second rotation of the two scalar components
\beq
\begin{pmatrix} h \\ H^0 \end{pmatrix} = \begin{pmatrix} \cos\alpha & \sin\alpha \\  -\sin\alpha & \cos\alpha \end{pmatrix} \begin{pmatrix} \rho_1 \\ \rho_2 \end{pmatrix}\,,
\eeq
where we can assume $0 \leq \alpha \leq \pi$ and $m_h < m_H$ without loss of generality.
The angle $\alpha$ describes the amount of mixing between the two scalar mass eigenstates. For simplicity, we take the limit $\alpha\to0$, which correspond to the heavy Higgs completely decoupling from the light counterpart and from the SM gauge bosons. Thus, $h$ is the SM Higgs.

Finally, the 2HDM Lagrangian is
\beq
{\cal L} ={\cal L}_{kin} + {\cal L}_Y + V(\Phi_1,\Phi_2)\,,
\eeq
where ${\cal L}_{kin}$ is the kinetic part for the two Higgs doublets, ${\cal L}_Y$ the Yukawa Lagrangian and $V(\Phi_1,\Phi_2)$ the scalar potential.

Firstly, the kinetic terms are
\beq
{\cal L}_{kin} =  D_\mu \Phi_1^\dagger D^\mu \Phi_1 +  \partial_\mu \Phi_2^\dagger \partial^\mu \Phi_2 \,,
\eeq
where we already imposed the decoupling limit $\alpha\to0$. The interaction of the light Higgs with the SM gauge bosons remain unaltered by this model and are given by the first term of ${\cal L}_{kin}$.

Secondly, the scalar potential $V(\Phi_1,\Phi_2)$ can be written in the general form
\beq
\begin{split}
V(\Phi_1,\Phi_2) &= m_1^2 \Phi_1^\dagger \Phi_1 + m_2^2 \Phi_2^\dagger \Phi_2 - \lp m_{12}^2  \Phi_1^\dagger \Phi_2 + {\rm h.c.} \rp + \frac{\beta_1}{2} \lp \Phi_1^\dagger \Phi_1 \rp^2   + \frac{\beta_2}{2} \lp \Phi_2^\dagger \Phi_2 \rp^2 \\
&+ \beta_3\lp \Phi_1^\dagger \Phi_1 \rp\lp \Phi_2^\dagger \Phi_2 \rp + \beta_4\lp \Phi_1^\dagger \Phi_2 \rp\lp \Phi_2^\dagger \Phi_1 \rp + \left[\frac{\beta_5}{2} \lp \Phi_1^\dagger \Phi_2 \rp^2 + {\rm h.c.}\right]\,.
\end{split}
\eeq
In the decoupling limit, all the mixing terms in the potential are taken to be zero, so $m_{12} = \beta_3 = \beta_4 = \beta_5 = 0$. Moreover, $m_1 = m_h = 125$ GeV and the quartic is $v^2 = -m_1^2/\beta_1$. Similarly, the mass term of the heavy Higgses are
\beq\label{eq:heavyH:mass}
m_H^2 \Phi_2^\dagger \Phi_2 = m_H^2 \lp |H^+|^2 + \frac{1}{2} |\eta|^2  + \frac{1}{2} |H^0|^2\rp\,, \qquad m_2 \equiv m_H\,.
\eeq
For simplicity, we neglect the quartic coupling of the heavy Higgs, that is $\beta_2 = 0$.

Finally, the Yukawa terms are
\beq
{\cal L}_Y = Y_{st}^u \bar q_s \tilde\Phi_1 u_t + Y_{st}^d \bar q_s \Phi_1 d_t + Y_{pr}^\ell \bar \ell_p \Phi_1 e_r + Y_{pr}^{',\ell}  \bar \ell_p  \Phi_2 e_r + Y_{st}^{',u}  \bar q_s \tilde \Phi_2 u_t + {\rm h. c.}\,,
\eeq
where $Y_{ij}^f$ and $Y_{ij}^{',f}$ are the $3
\times3$ Yukawa matrices, $\tilde \Phi_{1,2} = i \tau_2 \Phi^*_{1,2} $\,, and $d$ is the down-quark RH singlet. Here we have already assumed that $\Phi_2$ only couples to RH up-quarks and leptons, while $\Phi_1$ also couples to RH down-quarks. Finally, we take $Y_{ij}^{',u} = y_t' \delta_{i3} \delta_{j3} $ in the up-quark mass basis and $Y_{ij}^{',\ell} = y_l' \delta_{il} \delta_{jl} $ in the charged lepton mass basis, with $y_t'$ and $y_l'$ real, not to introduce new sources of CP violation. 

In terms of the doublets components and fermion mass eigenstates, the Yukawa Lagrangian reads
\beq\label{eq:L2HDM:Yukawa}
\begin{split}
{\cal L}_Y &\supset \lp v+h \rp \sum_f \frac{m_f}{v}  \bar f P_R  f  + \lp \frac{\eta + H^0}{\sqrt{2}} \rp \left[ y'_t \bar t P_R t + y'_l \bar l P_R  l \right] \\
&\quad \\
&+ H^+ \left[ y'_t V_{ti} \bar t P_L d_i + y'_l  \bar \nu_l P_R l \right] + {\rm h.c.}\,,
\end{split}
\eeq
where the elements of the $\Phi_1$ Yukawa matrices are written in the fermion mass basis using the tree-level matching condition as $Y_{ii}^f = m_f\sqrt{2}/v$, and $\nu_l$ denotes the $l$-flavored neutrino field.
With a small abuse of notation, now  $t$ and $d_i$ represent the top-quark and down-type quark (Dirac) fields in their mass basis, respectively, and $V_{ij}$ denote the CKM matrix elements.

\subsection{Phenomenology: full 2HDM vs SMEFT}

Here we describe the effect of the 2HDM on the two physical observables of interest, the muon anomalous magnetic moment and the signal strength of $h\to\mu^+\mu^-$ decay. In the full 2HDM calculation, these contributions consist of unresummed but finite loop diagram calculations. 
In order to compare them with the SMEFT result, we need to match the full theory to SMEFT at the UV matching scale $\Lambda$.
In the following we take $\Lambda = m_H$ without loss of generality. The tree level exchanges of neutral and charged heavy Higgs then generate the scalar four-fermion operator ${\cal O}_{4,\ell}$ in Eq.~\eqref{eq:operators:4and5} with the Wilson coefficient
\beq\label{eq:matching:tree}
{\hat \C_{4,\ell}} = \frac{y_t'y_\ell'}{m_H^2}\,.
\eeq
The latter is taken as the initial condition for the RGE in Appendix \ref{app:oneloop}. The subsequent running of the Wilson coefficients to the weak scale and matching to the low energy observables, as described in Section~\ref{sec:matchingbelowEW}, yields the desired result. The full 2HDM and SMEFT results are shown in Fig.~\ref{fig:twoloop:comparison} and Fig.~\ref{fig:THDM_HD} for $a_\mu$ and $h\to\mu^+\mu^-$ respectively.

One can in principle improve the matching condition in Eq.~\eqref{eq:matching:tree} by including one-loop effects for the other coefficients as initial conditions for the RGE, e.g. generating $\hat \C_{5,\ell}$ via box diagrams with one heavy Higgs and one gauge boson, or $\hat C_{1,\ell}$ via a heavy Higgs tadpole diagram with top quarks running in the loop. However, the effect of these is just to reduce the $\Lambda$ scale dependence of the matching away from $\Lambda=m_H$. Importantly, at one loop, they do not induce finite threshold effects as we have also checked explicitly. Consequently we do not consider them in the following.  

\subsubsection{$a_\mu$}
The additional ($\Phi_2$) Yukawa terms in Eq. \eqref{eq:L2HDM:Yukawa} will induce loop corrections to the lepton-photon vertex. It is well-known that in the 2HDM the two-loop Barr-Zee contributions to the anomalous magnetic moment generically dominate over one-loop effects~\cite{Ilisie_2015,Cherchiglia_2017,Botella_2020}. The chirality flip required for $a_\ell$ namely suppresses the latter, so that the former have a relative $(M/m_\ell)^2$ enhancement, where $M = m_t, m_W$ is a heavy particle in the loop. 

\begin{figure}[t]
	\centering
	\includegraphics[width=1\linewidth]{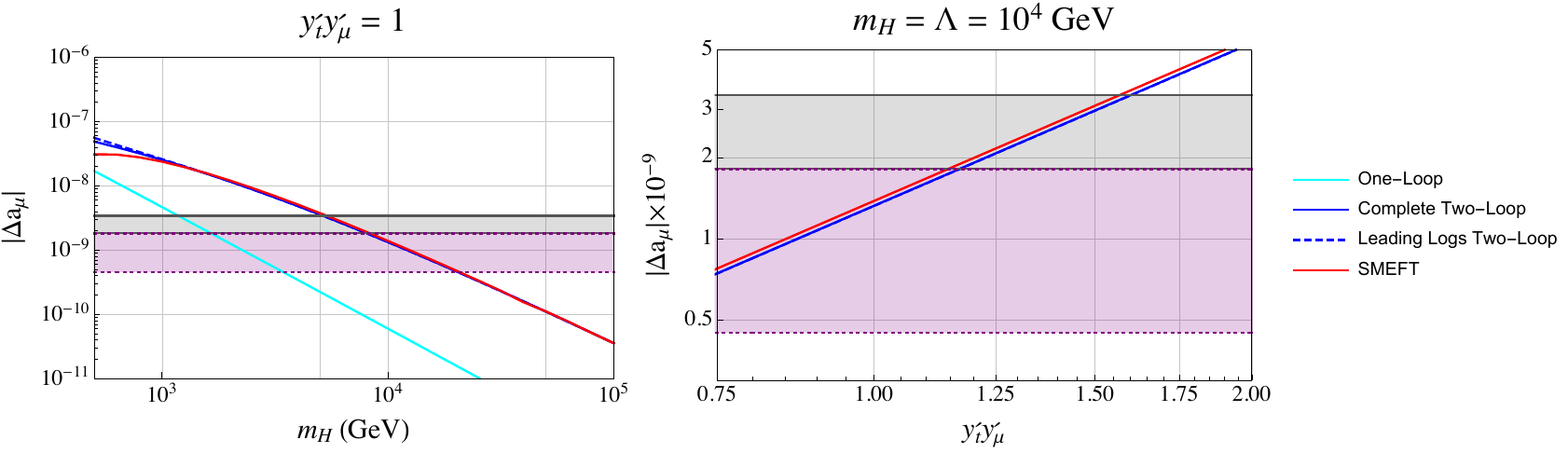} 
	\caption{\textbf{Left}: Total contribution to $\Delta a_\mu$ for different values of $m_H$, with fixed couplings to top and muon, compared to the preferred region by the muon anomalous magnetic moment, represented by the gray and purple bands (see Sec.~\ref{subsec:pheno:muon} for details). The cyan line shows the correction at one-loop order, while the blue full one includes the two-loop Barr-Zee diagrams. The blue dashed line is obtained by using only leading logarithm terms, see Appendix~\ref{app:leadlog}. The red line instead shows the SMEFT running and matching with Eq.~\eqref{eq:matching:tree} as the initial condition and $\Lambda = m_H$. 
		\textbf{Right}: Total contribution for different values of the product $y'_ty'_\mu$, at fixed $m_H = \Lambda = 10 $\,TeV. } \label{fig:twoloop:comparison}
\end{figure}

The complete one-and two-loop contributions to $a_\ell$ have been calculated for a generic 2HDM in Ref. \cite{Ilisie_2015}. As an example, in Fig.~\ref{fig:twoloop:comparison} left we show the results for $\ell=\mu$, where the cyan and blue lines represent the one-loop and two-loop results, respectively, as a function of the heavy Higgs mass $m_H$. The effect at two-loops is about one order of magnitude larger than the one-loop result for the mass range scanned. Fig.~\ref{fig:twoloop:comparison} right shows the same comparison when we fix $m_H = 10^4$ GeV and scan over the couplings. 

In the $m_H \gg m_t$ limit we expect the two-loop result to be dominated by logarithms of the form  
$\ln\lp m_t^2/m_H^2\rp$. Explicitly, the loop functions in Ref.~\cite{Ilisie_2015} can be expanded to keep only the leading terms in $\omega= m_t^2/m_H^2$, as shown in Appendix \ref{app:leadlog}. 
The blue dashed line in Fig.~\ref{fig:twoloop:comparison} shows this ``leading logarithm'' approximation as function of $m_H$ and $y'_t y'_\mu$ respectively. This approximation follows well the complete result all the way down to the weak scale. 

The SMEFT result takes as input only the matching in Eq.~\eqref{eq:matching:tree} as a function of the heavy Higgs mass and its couplings to leptons and up-quarks at $\Lambda=m_H$, RGE evolves it down to the weak scale and finally matches it to the relevant observables. Remarkably, as can be seen in Fig.~\ref{fig:twoloop:comparison} these ingredients are sufficient to reproduce the full theory result with very good accuracy. The excellent agreement can be partly understood by the fact that the full two-loop result is well approximated by its leading logarithms, which are the only terms effectively contained in RGE evolution of the tree-level or one-loop matched SMEFT. Only when $m_H\lesssim1$ TeV, the analytic contributions start to dominate, which cannot be reproduced by the RGE evolution of the effective theory.

\subsubsection{$h\to\mu^+\mu^-$}

In the $\beta = \alpha=0$ limit, there is no 2HDM tree level diagram entering in the Higgs decay. 
At one-loop however, a heavy neutral Higgs tad-pole contribution can avoid the lepton mass chiral suppression and yield a potentially important effect.  The details of such a calculation are given in Appendix~\ref{app:oneloopcalc:O1O4} with the result
\beq
\label{eq:c1thdm}
\delta c_{1\ell}  = N_c\frac{y'_\ell y'_t v^2}{4 \sqrt 2 m_H^2} \frac{y_t^3  }{(4\pi)^2}  {\rm ln}\lp\frac{m_t^2}{m_H^2}\rp\,.
\eeq

\begin{figure}[t]
	\centering
	\includegraphics[width=1\linewidth]{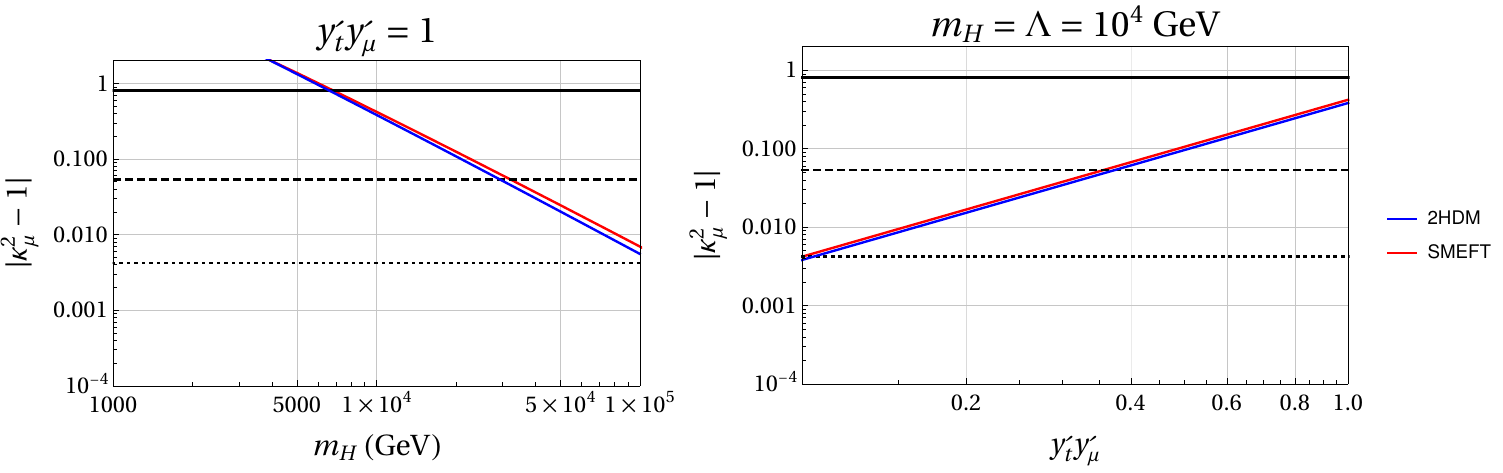} 
	\caption{\textbf{Left}: Shift to $h\to\mu^+\mu^-$ decay for different values of $m_H$, with fixed couplings to top and muon, compared to the present and future collider sensitivity, represented by the black solid (LHC), dashed (HL-LHC) and dotted (FCC) lines (see Sec.~\ref{subsec:pheno:muon} for details). Blue and red lines represent the full 2HDM contribution in Eq.~\eqref{eq:c1thdm} and the SMEFT running and matching, with Eq.~\eqref{eq:matching:tree} as initial condition, respectively.   \textbf{Right}: Total shift for different values of the product $y'_ty'_\mu$, with fixed $m_H = \Lambda =10$\,TeV. } \label{fig:THDM_HD}
\end{figure}

In Fig.~\ref{fig:THDM_HD} we compare the shift to $h\to\mu^+\mu^-$ decay obtained by generating $\delta c_{1\ell}$ in the full 2HDM (the blue line) and by matching the 2HDM to SMEFT at tree level, Eq.~\eqref{eq:matching:tree}, and consequently running to the weak scale (the red line). Similarly to the anomalous magnetic moment, the SMEFT result reproduces the full theory one to $5-10\%$ accuracy.

\section{Scalar LeptoQuarks}\label{sec:LeptoQuarks}
Here we briefly discuss a different UV completion via the introduction of LeptoQuarks (LQ). These particles can turn quarks into leptons (and viceversa) and can then provide a simple but compelling model. In particular, scalar LQ can generate four-fermion scalar and tensor operators already at tree level and contribute to the muon anomalous moment and Higgs decay to two muons. 

A comprehensive analysis of all possible allowed LQ models can be found in Ref.~\cite{Dor_ner_2016}, which we follow in the rest of this Section. There is a total of 12 different LQ realizations under the SM gauge group ${\cal G}_{SM} = SU(3)_C\times SU(2)_L\times U(1)_Y$. We focus on the scalars $S_1 = (\bar{\textbf{3}}, \textbf{1}, 1/3)$ and $R_2 = (\textbf{3}, \textbf{2}, 7/6)$ as the chirality structure of their interactions to fermions will lead to the operators in Eq.~\eqref{eq:operators:4and5}.

The interaction Lagrangians are respectively
\begin{align}
{\cal L}_{S_1} &\supset y_{1,ij}^{LL} \bar q^{C,ia} S_1 \epsilon^{ab}\ell^{j,b} + y_{1,ij}^{RR} \bar u^{C,i} S_1 \epsilon^{ab}e^{j} + {\rm h.c.}\,, \label{eq:Lagr:S1}\\ \nonumber\\
{\cal L}_{R_2} &\supset -y_{2,ij}^{RL} \bar u^i R_2^a \epsilon^{ab}\ell^{j,b} + y_{2,ij}^{LR} \bar e^i R_2^{a~*}q^{j,a} + {\rm h.c.}\,,\label{eq:Lagr:R2}
\end{align}
where we only kept the terms relevant for our discussion. Here $i,~j$ are generation indices, while $a,~b$ are $SU(2)$ indices. The superscripts of $y$ indicate the chirality properties of the LQ. In general, $y$ are arbitrary complex matrices in the generation space. 

\begin{figure}[t]
	\centering
	\includegraphics[width=1\linewidth]{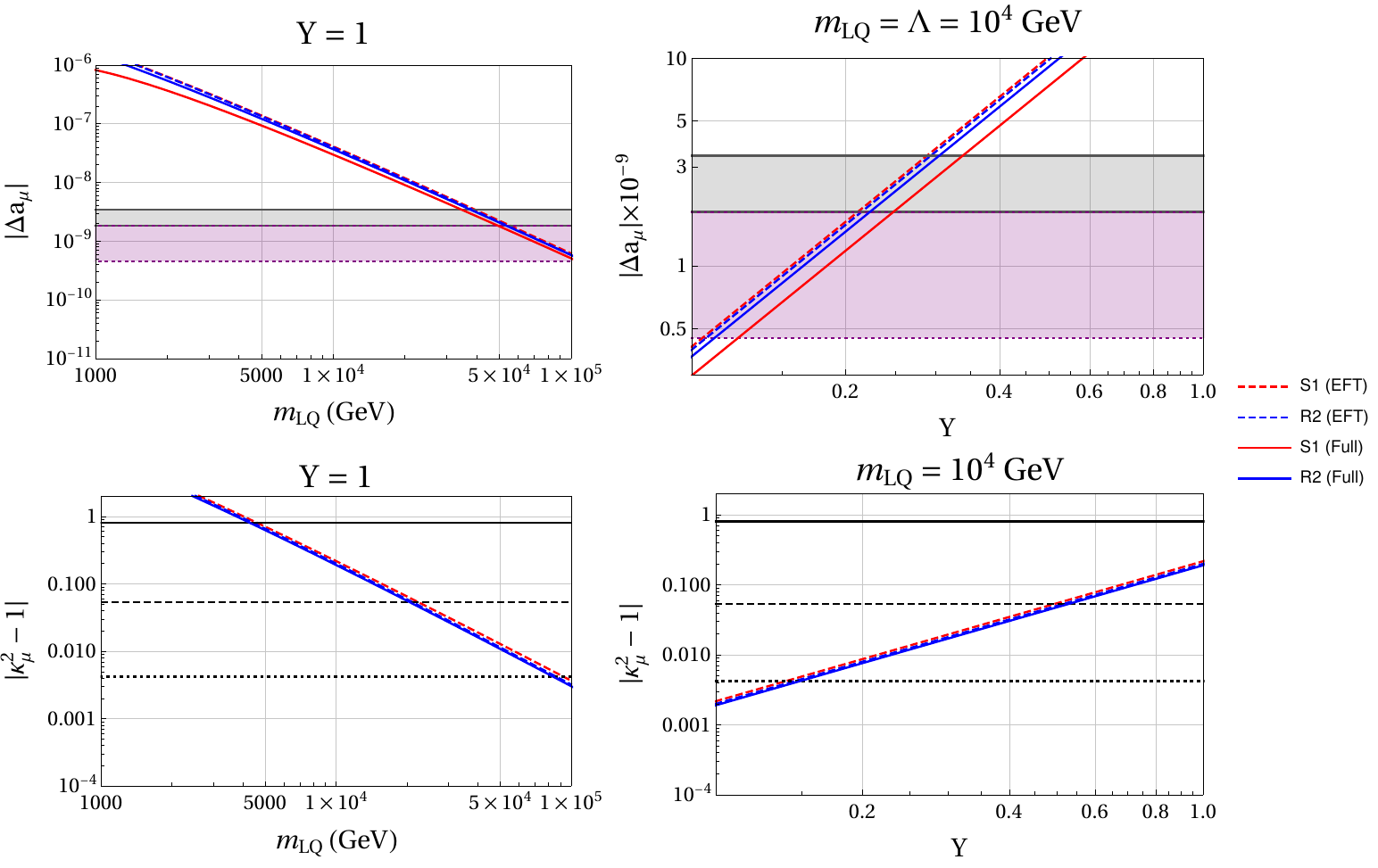} 
	\caption{\textbf{Top}: Shift to muon anomalous magnetic moment induced by LeptoQuarks for different values of $m_{LQ}$, with fixed couplings to muons and top quarks (left) and for different values of the coupling $Y$, with fixed LQ mass $m_{LQ} = \Lambda = 10^4$ GeV (right). The red and blue dashed lines show the result of matching LQ models to SMEFT operators using Eq.~\eqref{eq:LQ:SMEFTmatching} as initial condition and running to the weak scale for $S_1$ and $R_2$ case respectively. The red and blue solid lines are obtained using the results in Ref. \cite{crivellin2020correlating}. \textbf{Bottom}: Same for the shift to Higgs to two muons decay. The black solid, dashed and dotted lines represent lower bounds on $|\kappa_\mu^2 - 1|$ from LHC, HL-LHC and FCC respectively (see Sec.~\ref{subsec:pheno:muon} for details). Note that the solid lines overlap in these plots.} \label{fig:leptoquarks}
\end{figure}

The tree-level exchange of LQ directly generates four-fermion operators, which can be related to Eq. \eqref{eq:operators:4and5} by Fierz transformations. The relevant Lagrangian is \cite{Dor_ner_2016}
\beq
{\cal L} \supset -\frac{4 G_F}{\sqrt{2}} \left[ g_{ij,ks}^{LL} \Big( \bar q_L^i q_R^j \Big) \Big( \bar\ell_L^k \ell_R^s \Big) + h_{ij,ks}^{LL} \Big( \bar q_L^i \sigma_{\mu\nu} q_R^j \Big) \Big( \bar\ell_L^k \sigma^{\mu\nu} \ell_R^s \Big)\right]\,,
\eeq
where $i,j,k,s$ are again generational indices.
The two coefficients are easily related to the Wilson coefficients defined in Eq.~\eqref{eq:operators:4and5}. The matching conditions with the Lagrangians in Eqs.~\eqref{eq:Lagr:S1} and \eqref{eq:Lagr:R2} depend on the scalar LQ we consider. For simplicity, we take the LQ mass to be the same, $m_{S_1} = m_{R_2} = m_{LQ}$. We have
\beq\label{eq:LQ:matching}
\begin{split}
&S_1: g_{ij,ks}^{LL} = -4h_{ij,ks}^{LL} = \frac{v^2}{4 m_{LQ}^2} y_{1,js}^{RR} \lp y_1^{LL}\rp_{ik}^*\,,\\
&R_2: g_{ij,ks}^{LL} = 4h_{ij,ks}^{LL} = -\frac{v^2}{4 m_{LQ}^2} y_{2,jk}^{RL} \lp y_2^{LR}\rp_{si}^*\,.
\end{split}
\eeq
In the numerical analysis we are interested in the $i=j=3$ and $k=s=2$, that is in top quarks and muons. To simplify the notation, we suppress these indices and use the definition $Y\equiv y_{1,32}^{RR} \lp y_1^{LL}\rp_{32}^* = y_{2,32}^{RL} \lp y_2^{LR}\rp_{32}^*$. Notice that the only difference between the two LQ in the low energy basis is the relative sign of the two coefficients. With this, the tree level matching with the SMEFT operator reads trivially
\beq\label{eq:LQ:SMEFTmatching}
\hat\C_{4,\mu} = \mp 4 \hat\C_{5,\mu} = -\frac{4 G_F}{\sqrt{2}} g^{LL} = \frac{v^2 G_F}{\sqrt{2}m_{LQ}^2} Y\,,
\eeq
where the relative sign for $\hat\C_{5,\mu}$ depends on the LQ as in Eq.~\eqref{eq:LQ:matching}.

We can now use the latter  as initial condition to perform the SMEFT running to the weak scale. Results for the muon anomalous magnetic moment are shown as red and blue dashed lines for $S_1$ and $R_2$ case respectively in the top plots of Fig.~\ref{fig:leptoquarks}. The bottom plots show instead the shift in the Higgs decay signal strength. These EFT results are compared with the evaluations of $\delta a_\mu$ and of the ratio ${\rm Br}(h\to\mu^+\mu^-)/{\rm Br}_{{\rm SM}}(h\to\mu^+\mu^-)$ obtained in Ref. \cite{crivellin2020correlating}, shown in red and blue solid lines for $S_1$ and $R_2$ case respectively. Again, we see that the tree-level matched SMEFT analysis is able to reproduce accurately the full theory calculations. 

Note that in this study we have omitted possible additional (quartic) couplings of the LQ to the Higgs. By varying the size of these couplings, one can obtain significantly different results, as already shown in Ref.~\cite{crivellin2020correlating}. In SMEFT, these terms would induce genuine new one-loop matching contributions to the SMEFT operator basis and as such their effects cannot be reproduced with tree-level matching.


\section{Conclusions}\label{sec:conclusions}

In this paper we examined New Physics effects in charged lepton anomalous magnetic moments and Higgs decays to charged lepton pairs within the SMEFT framework. As the scale of NP is assumed to be well above the weak scale, we take into account the one-loop improved RGE evolution and mixing of the relevant operators. Just three dimension six chirality flipping operators including a Higgs and $SU(2)_L$ gauge bosons are sufficient for a tree-level analysis, while to form a closed set under the RGE we need to include four-fermion scalar and tensor operators. These can provide sizable corrections through top quark loops, avoiding typical chiral suppression from lepton masses. We run the SMEFT operators to the weak scale and match them to low-energy observables describing Higgs decay and anomalous magnetic moments. We derive the analytical formulae for dominant matching contributions at one-loop, which significantly reduce the associated EW matching scale dependence,  and perform a detailed numerical analysis of $\delta a_\ell$ and $h\to\ell^+\ell^-$ for each charged lepton, comparing several SMEFT benchmark scenarios to the current experimental results and future sensitivity projections. 

In the case of the electron, the SM predicted branching fraction of $h \to e^+ e^-$ cannot be reached with existing LHC data nor with the expected statistics of the HL-LHC upgrade. On the other hand, the existing discrepancy in $a_e$ can be addressed by the presence of a scalar four-fermion operator generated at a scale $\Lambda\gtrsim10$ TeV, which can induce the appropriate dipole operator and at the same time potentially impact  $h \to e^+ e^-$, measurements at the future planned FCC-ee experiments.  

In the case of the muon, the anomalous magnetic moment measurement currently offers a much better opportunity to probe NP than the recent LHC measurement of Higgs decay branching fraction to muon pairs. The scalar four-fermion operator can simultaneously solve the current discrepancy in $a_\mu$ and respect Higgs decay constraints, while requiring a relatively low NP scale, $\Lambda\sim{\cal O}(1-10)$ TeV. The HL-LHC upgrade will potentially be able to probe the associated effects in $h\to \mu^+ \mu^-$ and thus differentiate between heavy NP scenarios addressing the  $a_\mu$ discrepancy. 
However, should the $a_\mu$ discrepancy be significantly reduced, as indicated by the recent Lattice computation of the LO-HVP~\cite{Borsanyi:2020mff}, the lower limit on the associated NP scale would increase beyond $\Lambda\sim10$\,TeV, and the related effect in Higgs decays could no longer be probed at any currently considered future experimental facility, except possibly a dedicated muon collider. 

In the case of the tau, the branching fraction of $h \to \tau^+ \tau^-$ is already precisely measured at the LHC, agrees well with its SM predicted value, and thus severely constrains the size of possible NP effects in $a_\tau$. On the other hand, the current direct experimental reach for $a_\tau$ is rather limited and still well above the SM predictions. Consequently, significant improvements in $a_\tau$ sensitivity are required to make it a promising direction for NP searches at existing or currently planned experiments.

Going beyond the pure SMEFT analysis, we explore two examples of UV complete models that generate scalar and tensor four-fermion operators. Firstly, we build a Two Higgs Doublet Model in the "decoupling" limit, where one doublet works as the SM Higgs and the other as a heavy scalar coupling only to leptons and quarks. We use known results in the literature for one- and two-loop diagrams contributing to the anomalous magnetic moment and compare these with the SMEFT matching and running procedure. We find that the latter (which only involves tree-level calculations in the UV model) can reproduce the full theory two-loop calculations to remarkable accuracy for $m_H \gtrsim 1$\,TeV. In the process we have clarified the role of higher order contributions to the matching of UV models to SMEFT. Keeping the heavy Higgs couplings to top quarks and muons below ${\cal O}(1)$, we also show how a heavy Higgs with $m_H \lesssim10$\,TeV can solve the current discrepancy, with possibly the only observable effect at the (HL-)LHC appearing in $h\to \mu^+\mu^-$. Secondly, we introduce the scalar LeptoQuarks $S_1$ and $R_2$, which possess the right chiral structure to generate at tree-level the operators in Eq.~\eqref{eq:operators:4and5}. The SMEFT tree-level matching and one-loop RGE running for this model is straightforward and again reproduces well results obtained in the full theory, (see Fig.~\ref{fig:leptoquarks}). The analysis shows how both LQ considered can solve the discrepancy in $a_\mu$ for masses $m_{LQ}\lesssim 50$ TeV and perturbative couplings. The associated effect in the Higgs decays to muons is then expected to be close to the projected sensitivity of the FCC. To summarize, the two UV model examples showcase the power of employing SMEFT in phenomenological analyses of explicit NP models. The procedure of matching and running is in principle doable with any other UV model and an appropriate set of SMEFT operators, and allows to conveniently capture the dominant effects of higher order contributions of heavy NP on low energy phenomenology.


\section*{Acknowledgments}
We thank Peter Stoffer for useful discussions and clarifications regarding Ref.~\cite{Dekens_2019}, as well as Yotam Soreq and Paride Paradisi for for their comments on the recent determinations of the fine structure constant. The authors acknowledge the financial support from the Slovenian Research Agency (research core funding No. P1-0035).

\begin{appendix}

\section{RGE}
\label{app:oneloop}
Here we collect the one-loop RG equations from Refs.~\cite{Jenkins_2013,Jenkins_2014,Alonso_2014} describing the evolution and mixing of the operator basis in Eqs. \eqref{eq:operators:1}-\eqref{eq:operators:3} and \eqref{eq:operators:4and5}. We assume that all Yukawa couplings are zero, except for the top-quark Yukawa. We define
\beq
\dot \C_{i,\ell} \equiv 16\pi^2\mu\frac{d\C_{i,\ell}}{d\mu}\,,
\eeq
where $\mu$ in this case is the running renormalization scale. The equations read
\setlength{\jot}{10pt}
\begin{align}
\label{eq:RGEC1}
\dot\C_{1,\ell} &= \left[-\frac{27}{4} g_2^2 - 3 \lp 3 Y_\ell^2 + 3 Y_e^2 - 4 Y_\ell Y_e \rp g_1^2  + 3 N_c y_t^2  + 24 \lambda \right] \C_{1,\ell} + 4 N_c y_t \lp y_t^2 - \lambda \rp \C_{4,\ell} \notag\\
&- 3  \lp 4 g_1^2 g_2 Y_h \lp Y_e + Y_\ell \rp + 3 g_2^3 \rp \C_{2,\ell} -6  \lp 4 g_1^3 Y_h^2 \lp Y_e  + Y_\ell \rp + g_2^2 g_1Y_h \rp \C_{3,\ell}\,,    \\
\dot\C_{2,\ell} &= \left[ \lp 3 c_{F,2} - b_{0,2} \rp g_2^2 + \lp -3 Y_e^2 + 8 Y_e Y_\ell - 3 Y_\ell^2 \rp g_1^2  + N_c y_t^2 \right] \C_{2,\ell} \\ 
&+ g_1 g_2 \lp 3 Y_\ell - Y_e \rp \C_{3,\ell} - 2 g_2 N_c y_t \C_{5,\ell}\,,  \notag\\
\dot\C_{3,\ell} &= \left[ -3 c_{F,2} g_2^2 + \lp 3 Y_e^2 + 4 Y_e Y_\ell + 3 Y_\ell^2 - b_{0,1} \rp g_1^2 + N_c y_t^2 \right] \C_{3,\ell} \\
&+ 4 c_{F,2} g_1 g_2 \lp 3 Y_\ell - Y_e \rp \C_{2,\ell} + 4 g_1 N_c y_t \lp Y_u + Y_q \rp \C_{5,\ell}\,, \notag \\
\dot\C_{4,\ell} &= -\left[ 6 \lp Y_e^2 + Y_e \lp Y_u - Y_q \rp + Y_q Y_u \rp g_1^2 + 3 \lp N_c - \frac{1}{N_c} \rp g_3^2 + y_t^2 \lp 2 N_c + 1 \rp \right] \C_{4,\ell} \\
&- \left [ 24 \lp Y_q + Y_u \rp \lp 2 Y_e - Y_q + Y_u \rp g_1^2 - 18 g_2^2 \right] \C_{5,\ell}\,, \notag \\
\dot\C_{5,\ell} &= g_1 \lp Y_q + Y_u \rp y_t \C_{3,\ell} - \frac{3}{2} g_2 y_t \C_{2,\ell} + \Big[ 2 \lp Y_e^2 - Y_e Y_q + Y_e Y_u - 2 Y_q^2 + 5 Y_q Y_u - 2 Y_u^2 \rp g_1^2 \notag\\
&- 3 g_2^2 + \lp N_c - \frac{1}{N_c} \rp g_3^2 + y_t^2\Big] \C_{5,\ell} + \frac{1}{8} \left[ - 4 \lp Y_q + Y_u \rp \lp 2 Y_e - Y_q + Y_u \rp g_1^2 + 3 g_2^2 \right] \C_{4,\ell}\,.
\end{align}
Here $Y_i$ are the hypercharges of the fermions, $N_c = 3$  is the number of colors and $n_g = 3$ the number of active generations. We also used the definitions $c_{F,2} = 3/4$, $b_{0,1} = -1/6 - 20 n_g/9$ and $b_{0,2} = 43/6 - 4 n_g/3$.

To form a closed set of equations under RG, we must include the running of the muon Yukawa, the Higgs mass and quartic coupling. The equations are \cite{Mihaila_2012,Chetyrkin_2012}
\begin{align}
\dot y_\ell &= y_\ell \lp \frac{3}{4} y_\ell^2 + \frac{3}{2} \lp y_t^2 + y_\ell^2 \rp - \frac{9}{8} \lp g_1^2 + g_2^2 \rp \rp +  m_h^2 \lp 3 \hat \C_{1,\ell} - N_c y_t \hat \C_{4,\ell} \rp\,, \label{eq:oneloop:muonyuk}\\
\dot\lambda &= 2 \lp 12 \lambda^2  + 2 N_c y_t^2 \lambda - N_c y_t^4 + m_h^2 y_\ell\hat \C_{1,\ell} \rp\,,\label{eq:oneloop:quartic}\\
\dot m_h &= m_h \lp 6 \lambda  +N_c y_t^2 - \frac{9}{4} g_2^2 - \frac{3}{4} g_1^2 \rp\,.\label{eq:oneloop:massH}
\end{align} 

\section{Details on one-loop diagram calculations in SMEFT}
\label{app:oneloopcalc}


Here we present the details on the calculation of one-loop diagrams needed to perform the matching of SMEFT to the low energy observables below the weak scale in Eqs.~\eqref{eq:deltay:matching} and~\eqref{eq:matching:Cprime} as well as the full one-loop contribution to Higgs decay to two leptons in the 2HDM in Eq.~\eqref{eq:c1thdm}. The Feynman Rules for the EFT vertices have been obtained independently by following Ref.~\cite{paraskevas2018dirac} and using {\tt FeynRules} \cite{Alloul_2014}. As we use dimensional regularization in these calculations, we treat the terms including a $\g_5$ considering the scheme convention in Ref.~\cite{Dekens_2019}.
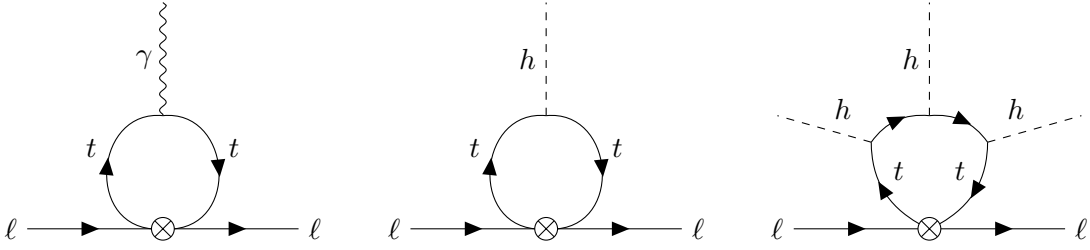
\begin{figure}[t]
	\centering
	\begin{tikzpicture}
	\begin{feynman}
	\vertex (a1) {\(\ell\)};
	\vertex[right=4cm of a1] (a3) {\(\ell\)};
	\vertex[right=2cm of a1, crossed dot] (a2) {};
	\vertex[above=of a2] (b1);
	\vertex[above=of b1] (b2);
	
	\diagram* {
		{[edges=fermion]
			(a1) -- (a2) -- (a3),
			(a2) --[half left, edge label=\(t\)](b1),
			(b1)--[half left, edge label=\(t\)](a2),
		},
		(b1) -- [boson, edge label=\(\gamma\)] (b2),
	};	
	\end{feynman}
	\end{tikzpicture}\hspace{.5cm}
	\begin{tikzpicture}
	\begin{feynman}
	\vertex (a1) {\(\ell\)};
	\vertex[right=4cm of a1] (a3) {\(\ell\)};
	\vertex[right=2cm of a1, crossed dot] (a2) {};
	\vertex[above=of a2] (b1);
	\vertex[above=of b1] (b2);
	
	\diagram* {
		{[edges=fermion]
			(a1) -- (a2) -- (a3),
			(a2) --[half left, edge label=\(t\)](b1),
			(b1)--[half left, edge label=\(t\)](a2),
		},
		(b1) -- [scalar, edge label=\(h\)] (b2),
	};	
	\end{feynman}
	\end{tikzpicture}\hspace{.5cm}
	\begin{tikzpicture}
	\begin{feynman}
	\vertex (a1) {\(\ell\)};
	\vertex[right=4cm of a1] (a3) {\(\ell\)};
	\vertex[right=2cm of a1, crossed dot] (a2) {};
	\vertex[above=3em of a2] (h1);
	\vertex[right=2em of h1] (h2);
	\vertex[left=2em of h1] (h3);
	\vertex[above=of a2] (b1);
	\vertex[above=of b1] (b2);
	\vertex[above=of a1] (b3);
	\vertex[above=of a3] (b4);
	
	\diagram* {
		{[edges=fermion]
			(a1) -- (a2) -- (a3),
			(a2) --[bend left, edge label'=\(t\)](h3),
			(h3)--[bend left](b1),
			(b1)--[bend left](h2),
			(h2)--[bend left, edge label'=\(t\)](a2),
		},
		(b1) -- [scalar, edge label=\(h\)] (b2),
		(h2) -- [scalar, edge label=\(h\)] (b4),
		(h3) -- [scalar, edge label'=\(h\)] (b3),
	};	
	\end{feynman}
	\end{tikzpicture}
	\caption{One-loop diagrams appearing in the EW matching of ${\cal O}_{5,\ell}$ to ${\cal Q}_{2,\ell}$ (left, Section~\ref{app:oneloopcalc:O5O1}), in the matching ${\cal O}_{4,\ell}$ to ${\cal Q}_{1,\ell}$ (center, Section~\ref{app:oneloopcalc:O4O2}) and in the 2HDM matching to $\mathcal O_{1,\ell}$ (right, in the full theory, the operator insertion should be replaced by a heavy Higgs propagator, see Section~\ref{app:oneloopcalc:O1O4} for details).}\label{fig:C4toploop}
\end{figure}


\subsection{One-Loop Matching of ${\cal O}_{4,\ell}$ to ${\cal Q}_{1,\ell}$}
\label{app:oneloopcalc:O4O2}

The operator ${\cal Q}_{1,\ell}$ can be generated by ${\cal O}_{4,\ell}$ by attaching one Higgs to the top loop, as shown in Fig.~\ref{fig:C4toploop} center. The external legs are leptons, with momenta $p_1$ (incoming) and $p_2$ (outgoing), so the Higgs momentum is $p_h = p_1 - p_2$. In the loop we only consider top quarks, as only the chirality flipping term will survive and will give a term proportional to the quark mass. The loop momenta are $l$ and $l'=l-p_h$.

The Feynamn rule for the effective vertex is 
\beq\label{eq:app:FR4}
FR_4 =i \hat\C_{4,\ell} \delta_{c_3,c_4} P_{1,2} P_{3,4}\,.
\eeq
In our case, ${1,2}$ are the two muons and ${3,4}$ are the two tops, both with $P_{i,j} = P_R = (1+\g_5)/2$. The indices $c_3,~c_4$ run over the top colors. We can now use this simple expression in the fermion loop trace.

The amplitude reads
\beq
\begin{split}
	{\cal M} &=\lp- i \frac{y_t}{\sqrt 2}\rp (i N_c \hat\C_{4,\ell}) \bar v_2 \int \frac{d^d l}{(2\pi)^d} \mu^{2\varepsilon} \left[ {\rm Tr} \lp\frac{ i(\slashed l + m_t) i(\slashed l' + m_t)P_R}{(l^2 - m_t^2)(l'^2 - m_t^2)} \rp\right] u_1\\
	&= \sqrt 2 y_t N_c \hat\C_{4,\ell} \bar v_2 \int_0^1 dx\int \frac{d^d k}{(2\pi)^d} \mu^{2\varepsilon}\left[ \frac{ k^2 + m_t^2}{(k^2 - \Delta)^2}\right] u_1\,.
\end{split}
\eeq
We further simplify this expression by taking $m_h \ll m_t$, or equivalently zero external momenta. Thus $p_h^2 = 0$ and $\Delta = m_t^2$. The integral in $dx$ is now trivial and the d-dimensional integral gives
\beq
{\cal M} = \sqrt 2 y_t N_c \hat\C_{4,\ell} \frac{im_t^2}{(4\pi)^{d/2}} \Gamma\lp 2-\frac{d}{2} \rp \lp \frac{1}{m_t^2} \rp^{2-d/2} \left[1 - \frac{d/2}{1-d/2} \right]{\cal Q}_{1,\ell}\,.
\eeq
Expanding $d\to4-2\varepsilon$ for $\varepsilon\to0$ and taking care of the pole, we have
\beq\label{eq:app:C2prime:oneloop}
\delta c_{1,\ell}(\mu_w)|_{1-\rm loop} = y_t N_c \hat\C_{4,\ell} \frac{3 m_t^2}{8 \sqrt 2 \pi^2} \left[ \frac{1}{3} + {\rm ln}\lp\frac{m_t^2}{\mu_w^2}\rp \right] \,.
\eeq
The overall factor of 3 in the latter equation (and in the tree level matching) is absorbed by renormalizing the lepton mass, giving the result eq.~\eqref{eq:deltay:matching}.

\subsection{One-Loop Matching of ${\cal O}_{5,\ell}$ to ${\cal Q}_{2,\ell}$}
\label{app:oneloopcalc:O5O1}
The four fermion operator ${\cal O}_{5,\ell}$ generates at one loop the dipole operator ${\cal Q}_{2,\ell}$. The diagram is represented in Fig.~\ref{fig:C4toploop} left, where a photon attaches to the quark in the loop.  We use the same momentum convention as before, with the replacement $q = p_h$ for the photon momentum. The Feynman rule reads 
\beq
FR_5 =\frac{i}{2} \hat\C_{5,\ell} \delta_{c_3,c_4} \lp \g^\mu\g^\nu P_{1,2} \g_\mu\g_\nu P_{3,4} - \g^\mu\g^\nu P_{3,4} \g_\nu\g_\mu P_{1,2} \rp\,,
\eeq
where the indices are the same convention as in eq.~\eqref{eq:app:FR4}. However, as shown in Ref.~\cite{Dekens_2019} (see their Section 4.2), we need to be careful to drop any linear term in $\varepsilon$. 

With these ingredients we can write the amplitude in dimensional regularization as
\beq
\begin{split}
{\cal M} &= iN_cQ_te \hat\C_{5,\ell} \bar u_2 \int \frac{d^d l}{(2\pi)^d} \mu^{2\varepsilon}\frac{i}{2} \g_\mu\g_\nu  \left[ {\rm Tr} \lp\frac{ i(\slashed l + m_t) \slashed \epsilon(q) i(\slashed l' + m_t)\g^\nu\g^\mu P_R}{(l^2 - m_t^2)(l'^2 - m_t^2)}\rp - {\rm Tr}(\mu\leftrightarrow \nu) \right] u_1\\
&= -N_c Q_te\hat\C_{5,\ell} \bar u_2 \int_0^1 dx\int \frac{d^d k}{(2\pi)^d} \mu^{2\varepsilon}\left[ \frac{4 m_t \slashed q \slashed \epsilon(q)}{(k^2 - \Delta)^2}\right]  u_1\,,
\end{split}
\eeq
where we defined $u_i \equiv u(p_i)$. $N_c = 3$ is the number of colors and $Q_t = 2/3$ is the top-quark electric charge. In the last step we used the decomposition in Feynman parameters and defined $k=l - (1-x)q$ and $\Delta=m^2 - q^2 x(x-1)$. Given the symmetry of the d-dimensional integral, terms with odd powers of $k$ vanish, thus the only piece remaining from the trace is the one proportional to $\slashed q \slashed \epsilon$. The integrals can be solved and we get
\beq
{\cal M} = \frac{ -4 i N_c m_t Q_te\hat\C_{5,\ell}}{(4\pi)^2} \bar u_2 \slashed q \slashed \epsilon  u_1 \lp \frac{1}{\hat\varepsilon} - {\rm ln}\lp\frac{m_t^2}{\mu^2}\rp\rp\,.
\eeq
In the previous equation we took the $q^2=0$ condition, so that $\Delta = m_t^2$ and the integration in $dx$ is trivial.

The $\slashed q \slashed \epsilon$ term can be manipulated to get the expression $\sigma_{\mu\nu} F^{\mu\nu}$. Indeed one can write 
\beq
\slashed q \slashed \epsilon = \g_\mu \g_\nu q^\mu \epsilon^\nu = \frac{2 i \sigma_{\mu\nu} + 2 \eta_{\mu\nu}}{2} \frac{q^\mu \epsilon^\nu - q^\nu \epsilon^\mu}{2} = - \frac{1}{2} \sigma_{\mu\nu} F^{\mu\nu} \to  - \frac{m_\ell}{2 e} {\cal Q}_{2,\ell}\,,
\eeq
where we used the commutator and anticommutator of $\g$ matrices, the antisymmetry of the tensor $q^\mu\epsilon^\nu$ and the product $q\cdot\epsilon(q)=0$ of photon momentum with its polarization vector. Also $i q^\mu = - \partial^\mu$.


The UV pole $1/\hat\varepsilon$ can be canceled by an appropriate counterterm, so that the contribution to the matching at the weak scale is 
\beq
 \delta c_{2,\ell}(\mu_w)|_{1-\rm loop} = \frac{N_c Q_t\hat\C_{5,\ell}(\mu_w) m_t m_\ell}{4 \pi^2}  {\rm ln}\lp\frac{m_t^2}{\mu_w^2} \rp \,.
\eeq

\subsection{2HDM Contribution to $h\to\ell^+\ell^-$ decay}
\label{app:oneloopcalc:O1O4}
In the 2HDM, shifts to the Higgs decay widths to leptons can be generated at one loop via mixing of the SM Higgs with the neutral heavy Higgses. The calculation is easier by considering instead the one-loop matching to SMEFT, see Fig.~\ref{fig:C4toploop} right, with the matching condition in eq.~\eqref{eq:matching:tree}. The pseudo-scalar Higgs contribution vanishes, while the charged Higgs can only enter at two-loop level. 
The lepton momenta are $p_1$ and $p_2$, while the external Higgs momenta are $k_1$, $k_2$ and $k_3$. Thus the four momenta in the loop are $l$, $l_1 = l + k_1$, $l_2 = l + k_1 + k_2$ and $l_3 = l + k_1 + k_2 + k_3$. There are 6 equal diagrams for this process.
The total amplitude then reads
\beq
\begin{split}
	{\cal M} &=  6 N_c \lp- i \frac{y_t}{\sqrt 2}\rp^3 \frac{i y'_\ell y'_t}{2 m_H^2} \bar v_2 \int \frac{d^d l}{(2\pi)^d} \mu^{2\varepsilon} \left[ {\rm Tr} \lp\frac{ i(\slashed l + m_t) i(\slashed l_1 + m_t) i(\slashed l_2 + m_t) i(\slashed l_3 + m_t) P_R}{(l^2 - m_t^2)(l_1^2 - m_t^2)(l_2^2 - m_t^2)(l_3^2 - m_t^2)} \rp\right] u_1\\
	&= - 3 N_c y_t^3  \frac{y'_\ell y'_t}{2\sqrt 2 m_H^2}  \bar v_2 \int_0^1 dx\int \frac{d^d k}{(2\pi)^d} \mu^{2\varepsilon}\left[ \frac{ k^4 + 6 k^2 m_t^2  + m_t^4}{(k^2 - m_t^2)^4}\right] u_1\,.
\end{split}
\eeq
where we have already taken zero external momenta. The integral in $d^dk$ is solved to
\beq
\label{eq:c12hdm}
\begin{split}
{\cal M} &= -  \frac{3 N_c y'_\ell y'_t}{m_H^2} \frac{iy_t^3  }{(4\pi)^\frac{d}{2}\Gamma(4)} \lp \frac{1}{m_t^2} \rp^{2 - \frac{d}{2}} \left[ \frac{d (d+2)}{4} \Gamma\lp 2 - \frac{d}{2}\rp - 3 d \Gamma\lp 3 - \frac{d}{2}\rp + \Gamma\lp 4 - \frac{d}{2}\rp \right] {\cal O}_{1,\ell} \\
&= \frac{3 N_c y_t^3}{16 \pi^2 } \frac{i y'_\ell y'_t  }{m_H^2}  \left[ {\rm ln}\lp\frac{m_t^2}{\mu^2} \rp - \frac{8}{3} \right] {\cal O}_{1,\ell}\,.
\end{split}
\eeq
In these last equations, a factor of $2 \sqrt 2$ has been re-absorbed into the operator definition of $ {\cal O}_{1,\ell}$, while the overall factor of 3 is absorbed by lepton mass renormalization as in Appendix~\ref{app:oneloopcalc:O4O2}. In the full 2HDM, the divergence in Eq.~\eqref{eq:c12hdm} is absorbed by the counterterm of the quartic  $Hhhh$ vertex, which we set to zero at the heavy Higgs mass scale ($\mu=m_H$).  From here (and using Eq.~\eqref{eq:deltay:matching}) we can read off the expression for $\delta c_{1,\ell}$ in Eq.~\eqref{eq:c1thdm}.  Conversely, when combined with Eq.~\eqref{eq:matching:tree} the logarithmic term on the SMEFT side (i.e. $\log (\mu^2/m^2_H)$ with $\mu \to \Lambda$) precisely matches the relevant $C_{4,\ell}$ RG coefficient in Eq.~\eqref{eq:RGEC1} and thus helps reduce the matching scale dependency away from $\Lambda=m_H$.

\section{Power expansion of two-loop Bar-Zee diagrams}
\label{app:leadlog}

Here we report the power expansion of the dominant terms in the Barr-Zee two-loop results for $\delta a_\mu$ within 2HDM presented in Ref.~\cite{Ilisie_2015}, which approximate well the full analytic results. Using the same notation as the reference, we have 
\beq\label{eq:app:leadlogs}
\begin{split}
	F_1(\omega) &\simeq \frac{\omega}{6} \lp -12 - \pi^2 - 6 \log\omega - 3 \log^2 \omega \rp\,, \\
	\tilde F_1 (\omega) &\simeq \frac{\omega}{6} \lp \pi^2 + 3 \log^2 \omega \rp\,,
\end{split}
\eeq
for the dominant neutral Higgs diagrams, and
\beq\label{eq:app:leadlogsG}
\begin{split}
	\int_0^1 dx \lp x Q_t + (1-x) Q_b \rp x (1-x) G (\omega,0) &\simeq \frac{1}{36} \lp -87 - 16 \pi^2 -78 \log\omega - 24 \log^2 \omega \rp \\
	&+ \frac{\omega}{9} \lp 21 + 8 \pi^2 -3 \log\omega + 12 \log^2 \omega \rp\,,
\end{split}
\eeq
for the dominant charged Higgs diagram. Here $Q_b = -1/3$ is the bottom quark electric charge and we neglected terms proportional to $m_b^2$, i.e. the bottom mass. The value of $\delta a_\mu$ obtained with this approximation is shown in Fig.~\ref{fig:twoloop:comparison}.

\end{appendix}
\bibliographystyle{h-physrev}
\bibliography{../gm2}

\end{document}